\PassOptionsToPackage{unicode}{hyperref}
\PassOptionsToPackage{naturalnames}{hyperref}
\documentclass[twocolumn]{aastex631}

\newcommand{\dmfrb}{\ensuremath{\rm DM_{FRB}}}
\newcommand{\dmigrm}{\ensuremath{\rm DM_{IGrM}}}

\newcommand{\dmunits}{\ensuremath{{\rm pc \, cm^{-3}}}}

\def\ra#1#2#3{#1$^{\rm h}$#2$^{\rm m}$#3$^{\rm s}$}
\def\dec#1#2#3{$#1^\circ#2'#3''$}

\usepackage{CJK}
\usepackage{amsmath}
\shorttitle{FRB\,20220610A Host}
\shortauthors{Gordon et al.}
\graphicspath{{./}{figures/}}

\begin{document}
\begin{CJK*}{UTF8}{gbsn}

\title{A Fast Radio Burst in a Compact Galaxy Group at $z \sim 1$}

\correspondingauthor{Alexa C. Gordon}
\email{alexagordon2026@u.northwestern.edu}

\newcommand{\NU}{\affiliation{Center for Interdisciplinary Exploration and Research in Astrophysics (CIERA) and Department of Physics and Astronomy, Northwestern University, Evanston, IL 60208, USA}}

\newcommand{\STSCI}{\affiliation{Space Telescope Science Institute, 3700 San Martin Drive, Baltimore, MD 21218, USA}}

\newcommand{\JHU}{\affiliation{Department of Physics and Astronomy, Johns Hopkins University, Baltimore, MD 21218, USA}}

\newcommand{\UCSC}{\affiliation{Department of Astronomy and Astrophysics, University of California, Santa Cruz, CA 95064, USA}}

\newcommand{\PUCV}{\affiliation{Instituto de F\'isica, Pontificia Universidad Cat\'olica de Valpara\'iso, Casilla 4059, Valpara\'iso, Chile}}

\newcommand{\IPMU}{\affiliation{Kavli Institute for the Physics and Mathematics of the Universe (Kavli IPMU), 5-1-5 Kashiwanoha, Kashiwa, 277-8583, Japan}}

\newcommand{\Swin}{\affiliation{Centre for Astrophysics and Supercomputing, Swinburne University of Technology, Hawthorn, VIC 3122, Australia}}

\newcommand{\MQ}{\affiliation{School of Mathematical and Physical Sciences, Macquarie University, NSW 2109, Australia}}

\newcommand{\ASTRMQ}{\affiliation{Astrophysics and Space Technologies Research Centre, Macquarie University, Sydney, NSW 2109, Australia}}

\newcommand{\CSIRO}{\affiliation{Australia Telescope National Facility, CSIRO Space and Astronomy, P.O. Box 76, Epping, NSW 1710, Australia}}

\newcommand{\NAOJ}{\affiliation{Division of Science, National Astronomical Observatory of Japan,2-21-1 Osawa, Mitaka, Tokyo 181-8588, Japan}}

\newcommand{\ARC}{\affiliation{ARC Centre of Excellence for All-Sky Astrophysics in 3 Dimensions (ASTRO 3D), Australia}}

\newcommand{\CMU}{\affiliation{McWilliams Center for Cosmology, Carnegie Mellon University, 5000 Forbes Avenue, Pittsburgh, PA 15213, USA}}

\newcommand{\MMO}{\affiliation{Maria Mitchell Observatory, 3 Vestal Street, Nantucket, MA 02554, USA}}

\newcommand{\ICRAR}{\affiliation{International Centre for Radio Astronomy Research (ICRAR), Curtin University, Bentley, WA 6102, Australia}}
\author[0000-0002-5025-4645]{Alexa~C.~Gordon}
\NU

\author[0000-0002-7374-935X]{Wen-fai~Fong}
\NU

\author[0000-0003-3801-1496]{Sunil~Simha}
\UCSC

\author[0000-0002-9363-8606]{Yuxin~Dong~(董雨欣)} 
\NU

\author[0000-0002-5740-7747]{Charles~D.~Kilpatrick}
\NU

\author[0000-0001-9434-3837]{Adam~T.~Deller}
\Swin

\author[0000-0003-4501-8100]{Stuart~D.~Ryder}
\MQ
\ASTRMQ

\author[0000-0003-0307-9984]{Tarraneh~Eftekhari}\thanks{NHFP Einstein Fellow}
\NU

\author[0000-0002-5067-8894]{Marcin~Glowacki}
\ICRAR

\author[0000-0003-1483-0147]{Lachlan~Marnoch}
\MQ
\CSIRO
\ASTRMQ
\ARC

\author[0000-0002-2849-6955]{August~R.~Muller}
\MMO

\author[0000-0002-2028-9329]{Anya~E.~Nugent}
\NU

\author[0000-0002-6011-0530 ]{Antonella~Palmese}
\CMU

\author[0000-0002-7738-6875]{J.~Xavier~Prochaska}
\UCSC
\IPMU
\NAOJ

\author[0000-0002-9946-4731]{Marc~Rafelski}
\STSCI
\JHU

\author[0000-0002-7285-6348]{Ryan~M.~Shannon}
\Swin

\author[0000-0002-1883-4252]{Nicolas~Tejos}
\PUCV



\begin{abstract} \label{sec:abstract}
FRB\,20220610A is a high-redshift Fast Radio Burst (FRB) that has not been observed to repeat. Here, we present rest-frame UV and optical \textit{Hubble Space Telescope} observations of the field of FRB\,20220610A. The imaging reveals seven extended sources, one of which we identify as the most likely host galaxy with a spectroscopic redshift of $z=1.017$. We spectroscopically confirm at least three additional sources to be at the same redshift, and identify the system as a compact galaxy group with possible signs of interaction among group members. We determine the host of FRB\,20220610A to be a star-forming galaxy with stellar mass of $\approx10^{9.7}\,M_{\odot}$, mass-weighted age of $\approx2.6$~Gyr, and star formation rate (integrated over the last 100~Myr) of $\approx1.7$~M$_{\odot}$~yr$^{-1}$. These host properties are commensurate with the star-forming field galaxy population at $z\sim 1$ and trace their properties analogously to the population of low-$z$ FRB hosts. Based on estimates of the total stellar mass of the galaxy group, we calculate a fiducial contribution to the observed Dispersion Measure (DM) from the intragroup medium of $\approx 110-220$~\dmunits\ (rest-frame). This leaves a significant excess of $500^{+272}_{-109}$~\dmunits\ (in the observer frame), with additional sources of DM possibly originating from the circumburst environment, host galaxy interstellar medium, and/or foreground structures along the line of sight. Given the low occurrence rates of galaxies in compact groups, the discovery of an FRB in such a group demonstrates a rare and novel environment in which FRBs can occur. 
\end{abstract}



\section{Introduction} \label{sec:intro}

Fast radio bursts (FRBs) are bright ($\approx10$~mJy--$100$~Jy; \citealt{Ravi19,Hashimoto22,Zhang23}), millisecond-duration radio bursts discovered at MHz--GHz frequencies. Despite hundreds of FRBs detected to date (e.g., \citealt{CHIME-catalog,Petroff+22}), many open questions remain, such as the physical distinction between FRBs that have been observed to repeat (``repeaters''; \citealt{Spitler+16,chime-repeater}) and those that have not (apparent ``non-repeaters''; \citealt{CHIME-catalog,Shannon+18}). Additionally, the progenitor systems and emission mechanisms for FRBs remain poorly understood, though their bright, coherent radio emission generally limits progenitor theories to radiation from compact objects (see \citealt{Platts+19} for a comprehensive review). Alongside studies of FRB properties, a promising avenue to decipher their progenitors is through detailed studies of their host galaxy environments to understand their stellar origins \citep[analogous to host studies of transients observed at other wavelengths such as supernovae and gamma-ray bursts, e.g.,][]{Bloom02,Sullivan06,Suzuki12,Schulze21,Nugent22}.  However, the majority of FRBs have thus far been localized to $\sim$ tens of arcminutes (mainly by the Canadian Hydrogen Intensity Mapping Experiment; \citealt{CHIME-catalog}), making routine identification of their host galaxies challenging.

Improvements in FRB experiments have enabled (sub-)arcsecond precision localizations \citep{Chatterjee+17,Law18,Bannister+19,Marcote_etal_2020,Law+23,Ravi23}. This degree of precision allows for robust associations to host galaxies \citep{Eftekhari2017}, allowing for identification of their host stellar population properties as observed in the rest-frame ultraviolet, optical, and infrared (e.g., \citealt{Tendulkar+17,Bannister+19,Prochaska+19,Ravi+19,Heintz+20,Marcote_etal_2020,Fong+21,Bhandari+22,Ravi22b,Bhandari_210117,Ravi22b,Niu+22,Ibik+23,Sharma_etal_2023,Law+23,Bhardwaj+23,Eftekhari2023}), morphologies \citep{Chittidi21,Mannings+21,Tendulkar+21,Xu+22,Dong+23}, and galactocentric offsets \citep{Heintz+20,Mannings+21,Bhandari+22}. The wealth of information uniquely enabled by host associations can provide strong constraints on FRB progenitor models via precise redshifts (therefore, rest-frame burst properties such as energetics) and properties of the local and global environments. Indeed, the first such host population studies found that FRBs occur in galaxies spanning a wide range of stellar population properties (e.g., stellar mass, stellar population ages, star formation rates, and metallicities). The majority are actively star-forming, with properties consistent with the field galaxy population at comparable redshifts \citep{Heintz+20,Bhandari+22,Gordon+23,Law+23}. Notably, two FRBs have been localized to galaxy clusters \citep{Connor_etal_2023,Sharma_etal_2023}, large collections of galaxies which generally have a  higher fraction of redder and quiescent galaxies than the field \citep{Balogh+98,Balogh+04}, and another has been associated with a group of WISE$\times$SCOS galaxies \citep{Rafiei_Ravandi_etal_2023}. All three of these FRBs in dense large-scale environments have been constrained to $z \lesssim 0.3$. While these hosts compose a small fraction of the total population, they highlight the diversity of FRB host environments and set a precedent for the association of FRBs to large-scale galactic structures.

Of the few dozen FRBs with redshifts determined from their host galaxies, the large majority reside at relatively low redshifts of $z \lesssim 0.5$ \citep{Ravi+19,Bhandari+22,Gordon+23,Ibik+23,LeeWaddell+23,Law+23,Panther+23,Ravi23}, primarily due to the sensitivity of current FRB discovery experiments \citep[e.g.,][]{Chime+18,Ravi23}. Thus, the population of FRB environments closer to the peak of cosmic star formation ($z\approx 2$; \citealt{md14}) is virtually uncharted. FRBs at $z\gtrsim 1$ are expected to be influential for several reasons. First, if FRB progenitors are dominated by a population with a prompt timescale relative to star formation (as some studies suggest, \citealt{James+22a,Gordon+23}, though c.f., \citealt{Zhang_and_Zhang19,Law+23}), the overall FRB rate should increase toward $z\approx2$. Pertinently, as the fraction of star-forming galaxies increases with redshift \citep{Whitaker12}, the localization of FRBs to quiescent galaxies, even at these redshifts, would indicate a significant contribution from a delayed (relative to star formation) progenitor or formation channel. Second, FRB sightlines at $z\gtrsim 1$ intersect longer stretches of the cosmic web, impacting the observed Dispersion Measure (DM), the integrated column density of electrons along the line of sight. Assuming FRBs reside in similar environments at $z \sim 1$ as they do at $z \lesssim 0.5$, the contribution of the host galaxy to the observed overall DM is expected to be smaller on average (as $(1+z)^{-1}$), thus serving as cleaner probes of the baryonic content in the Universe. As new facilities and sensitivity improvements to existing experiments continue to come online \citep[e.g.,][]{Ravi23,CHIME-Tone}, the prospects for detecting and localizing FRBs to higher redshift will continue to improve.

The first such event, FRB\,20220610A, was discovered in June 2022 with the Australian Square Kilometre Array Pathfinder (ASKAP; \citealt{ASKAP}) by the Commensal Real-Time ASKAP Fast-Transients collaboration (CRAFT; \citealt{mbb+10,Ryder23}). To date, FRB\,20220610A is not known to repeat. Follow-up imaging on the 8.2~m Very Large Telescope (VLT) revealed a ``complex host galaxy system'' with three blended clumps. Spectroscopy of these clumps indicated a redshift of $z=1.016\pm0.002$, establishing FRB\,20220610A as the highest confirmed redshift FRB to date. The redshift also enabled a measurement of the FRB burst energy; at $\sim$2$\times$10$^{42}$~erg, it was $\approx 4\times$ greater than any other known FRB. The observed DM of FRB\,20220610A was commensurately large, at 1458~\dmunits\ \citep{Ryder23}. However, in some cases, the inferred redshift from the Macquart relation is higher than the observed redshift, resulting in an ``excess'' of DM \citep{Niu+22,Simha+23,Connor_etal_2023}. Despite its high redshift, FRB\,20220610A had a DM excess of $\approx650$~\dmunits, which \citet{Ryder23} attributed to the host galaxy. \citet{Ryder23} posited the three clumps identified in the imaging may either represent a single galaxy or a system comprised of several interacting galaxies. Due to the limited spatial resolution and sensitivity of the ground-based imaging, they were unable to distinguish between these scenarios. 

In this paper, we present \textit{Hubble Space Telescope (HST)} observations of the host galaxy system of FRB\,20220610A and its immediate environment, identifying the host galaxy for the first time (Section~\ref{sec:obs}). In Section~\ref{sec:results}, we detail the stellar population properties, morphology, and light distributions of the \textit{HST} imaging of the host; we additionally characterize the morphologies of the surrounding galaxies. We discuss the implications of these results in Section~\ref{sec:disc}, including commenting on the nature of the DM excess, and summarize and conclude our analysis in Section~\ref{sec:conc}. For all relevant calculations in this work, we assume WMAP9 cosmology \citep{WMAP9}.

\section{Observations and Data Reduction} \label{sec:obs}
\subsection{HST Observations} \label{sec:hst-obs}

The host galaxy of FRB\,20220610A was observed with the \textit{HST} Wide Field Camera 3 (WFC3) using the ultraviolet-visual (UVIS) and infrared (IR) channels under Program GO-17277 (PI: Gordon). We obtained two orbits in each of the F606W and F160W filters to probe the rest-frame ultraviolet and rest-frame optical light, respectively. These observations total 5055\,s in the F606W filter and 4824\,s in the F160W filter. We detail the observations in Table~\ref{tab:Obs-log}.

To avoid cosmic ray contamination and obtain high-quality pixel sampling to subsample the point spread function (PSF), we used a custom 6- and 8-point dither pattern for the F606W and F160W images, as described in WFC3/ISR 2020-07 \citep{Anderson2020} and WFC3/ISR 2019-05 \citep{Mack2019}, respectively. We increased the size of the WFC3/UVIS pattern relative to the default pattern to dither over the fixed pattern noise in the dark frames (see \citealt{Rafelski_etal_2015}). By obtaining three exposures per orbit in F606W, we ensured the background was high enough (above 20e$^{-}$ per pixel) to avoid using post-flash despite the degraded charge transfer efficiency (CTE). We additionally placed the target on chip 2 close to the readout region to minimize the effects of CTE. For the F160W observations, we used SPAR50 with an NSAMP of 13 to minimize persistence while maximizing sensitivity. We also tied the orientations of the observations together to ensure sufficient common background sources for alignment.

We retrieved the data from the Barbara A. Mikulski Archive for Space Telescopes (MAST)\footnote{\url{https://mast.stsci.edu/}}. We apply cosmic-ray corrections and combine the raw images using \texttt{AstroDrizzle} (v.3.5.1) in \textsc{drizzlepac} (v.3.5.1) \citep{drizzlepac}. We perform relative astrometry to the VLT $R_{\rm special}$-band image \citep{Ryder23}, which itself is tied to Gaia DR3 \citep{Gaia,Gaia2}, using \texttt{TweakReg} (v3.5.1) and {\tt Source Extractor} \citep{source_extractor}. For F606W, we used 15 common sources to perform this alignment, resulting in an astrometric uncertainty of $0\farcs024$. For F160W, we used 22 sources in common, giving an astrometric uncertainty of $0\farcs054$. Finally, we re-sampled the images to improve resolution using a \texttt{pixfrac} of 0.8 and a \texttt{pixscale} of 0.03 ($0\farcs03$) and 0.06 ($0\farcs06$) for the F606W and F160W images, respectively. The host galaxy, as well as several other surrounding sources, are clearly detected in both filters. We further describe these in Section~\ref{sec:system}. We present the F606W and F160W imaging in Figure~\ref{fig:imaging}.

\subsection{Literature Data} \label{sec:lit-data}

We complement our {\it HST} data with those obtained in \citet{Ryder23}. Specifically, we use the $g$- and $R_{\rm special}$-band imaging taken with the FOcal Reducer and low dispersion Spectrograph 2 mounted on the VLT (FORS2; PI: Shannon, Program 108.21ZF.001; \citealt{FORS2}); $J$- and $K_{\rm s}$-band imaging with the High Acuity Wide-field K-band Imager on VLT (HAWK-I; PI Shannon, Programs 108.21ZF.006, 108.21ZF.005; \citealt{HAWKI}); and $V$-, $I$-, and $Z$-band imaging with the Deep Imaging Multi-Object Spectrograph on the 10m Keck II Telescope (DEIMOS; PI: Weisz, Program U028; \citealt{DEIMOS}). The details of the data reduction are in \citet{Ryder23}, and further details of these observations are listed in Table~\ref{tab:Obs-log}. Additionally, we use the two VLT/X-Shooter spectra obtained by \citet{Ryder23} (PI Macquart, Program 105.204W.004; \citealt{X-shooter}), further detailed in Section~\ref{sec:system} and Appendix~\ref{app:emission-lines}.

\begin{deluxetable*}{cccccccc}
\tablewidth{0pc}
\tablecaption{Observation Log of the Host and Surrounding Galaxies of FRB\,20220610A
\label{tab:Obs-log}}
\tablehead{
\colhead{Date} &
\colhead{Facility/Instrument} &
\colhead{Filter} &
\colhead{Galaxy} & 
\colhead{Magnitude} &
\colhead{A$_{\lambda}$} &
\colhead{Exp. Time} &
\colhead{Program ID} \\ 
\colhead{} &
\colhead{} &
\colhead{} &
\colhead{} &
\colhead{(AB~mag)} &
\colhead{(mag)} &
\colhead{(s)} &
\colhead{}
}
\startdata
2022 Oct 2-4 UTC & VLT/FORS2 & $g$ & 1 & 24.66$\pm$0.36 & 0.060 & 6000 & 108.21ZF.001 \\
 & & & 6 & $>$25.9 & & & \\
2022 Sept 24 UTC & Keck/DEIMOS & $V$ & 1 & 24.61$\pm$0.24 & 0.054 & 1050 & U028 \\
 & & & 6 & $>$24.3 & & & \\
2023 June 10 UTC & \textit{HST}/WFC3-UVIS & F606W & 1 & 24.89$\pm$0.04 & 0.044 & 5055 & GO 17277 \\
 & & & 2 & 25.34$\pm$0.04 & & & \\
 & & & 3 & 27.83$\pm$0.28 & & & \\
 & & & 4 & 25.65$\pm$0.29 & & & \\
 & & & 5 & 26.65$\pm$0.29 & & & \\
 & & & 6 & 25.98$\pm$0.06 & & & \\
 & & & 7 & 27.40$\pm$0.17 & & & \\
2022 July 1 UTC & VLT/FORS2 & $R_{\rm special}$ & 1 & 24.76$\pm$0.25 & 0.038 & 2000 & 108.21ZF.001 \\
 & & & 6 & 24.32$\pm$0.39 & & & \\
2022 Sept 24 UTC & Keck/DEIMOS & $I$ & 1, 6 & $>$24.0 & 0.025 & 1050 & U028 \\
2022 Sept 24 UTC & Keck/DEIMOS & $Z$ & 1, 6 & $>$24.5 & 0.021 & 1050 & U028 \\
2022 Sept 29 UTC & VLT/HAWK-I & $J$ & 1 & 23.02$\pm$0.12 & 0.012 & 150 & 108.21ZF.006 \\
 & & & 6 & 24.32$\pm$0.22 & & & \\
2023 April 26 UTC & \textit{HST}/WFC3-IR & F160W & 1 & 23.11$\pm$0.01 & 0.008 & 4824 & GO 17277 \\
 & & & 6 & 24.75$\pm$0.13 & & & \\
2022 July 24 UTC & VLT/HAWK-I & $K_{\rm s}$ & 1 & 22.85$\pm$0.18 & 0.004 & 150 & 108.21ZF.005 \\
 & & & 6 & $>$24.0 & & & \\
\enddata
\tablecomments{
The details of observations used in this work including observation date, telescope and instrument, filter, exposure time, and program ID. The apparent magnitudes are corrected for Galactic extinction, A$_\lambda$, following the \cite{Fitzpatrick:2007} extinction law. \\}
\end{deluxetable*}

\section{Results} \label{sec:results}
\subsection{Host Association and Identification of Multiple Galaxies at $z$$\approx$$1$}\label{sec:system}

\begin{figure*}
    \centering   
    \includegraphics[width=0.49\textwidth]{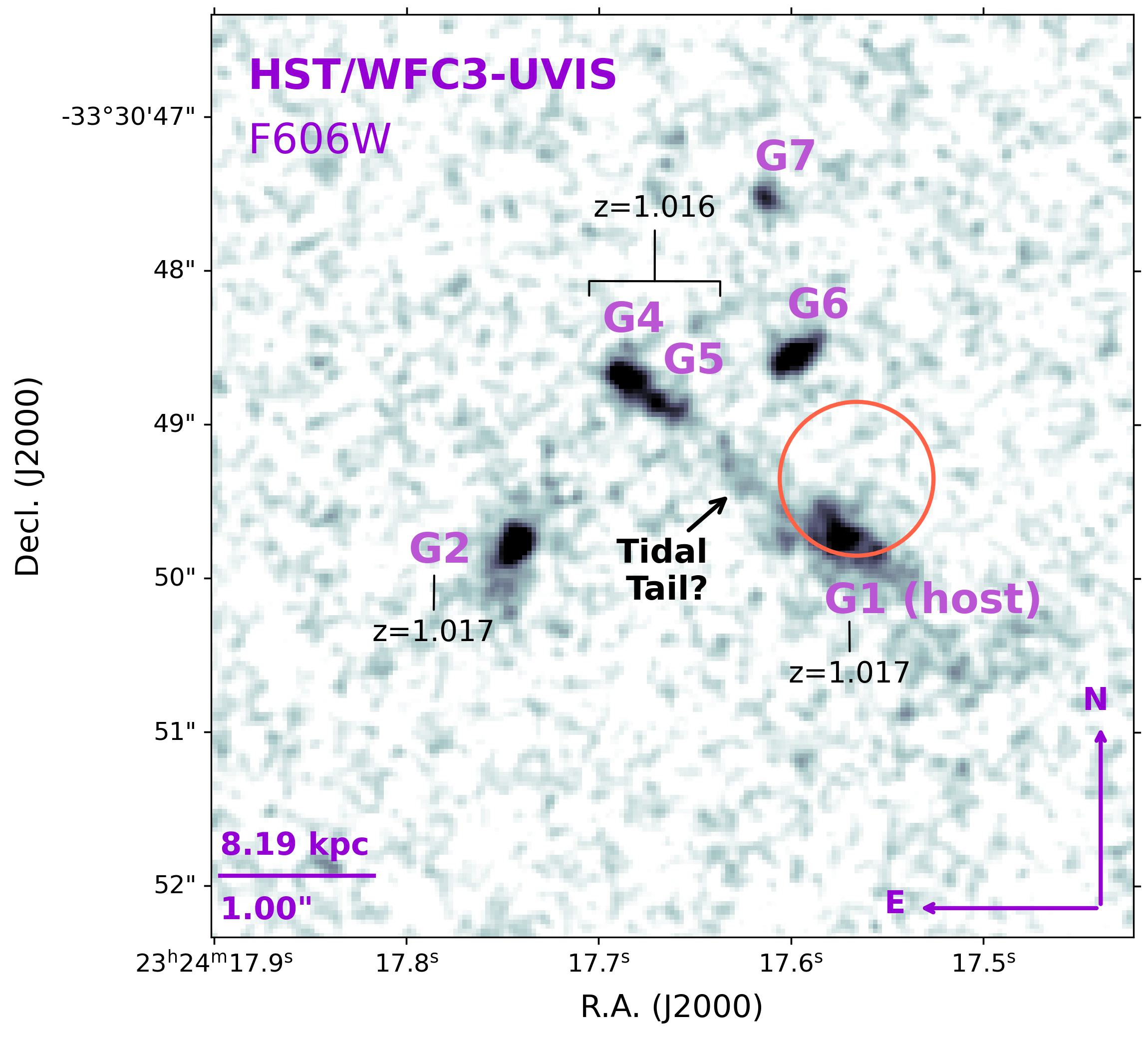}
    \includegraphics[width=0.49\textwidth]{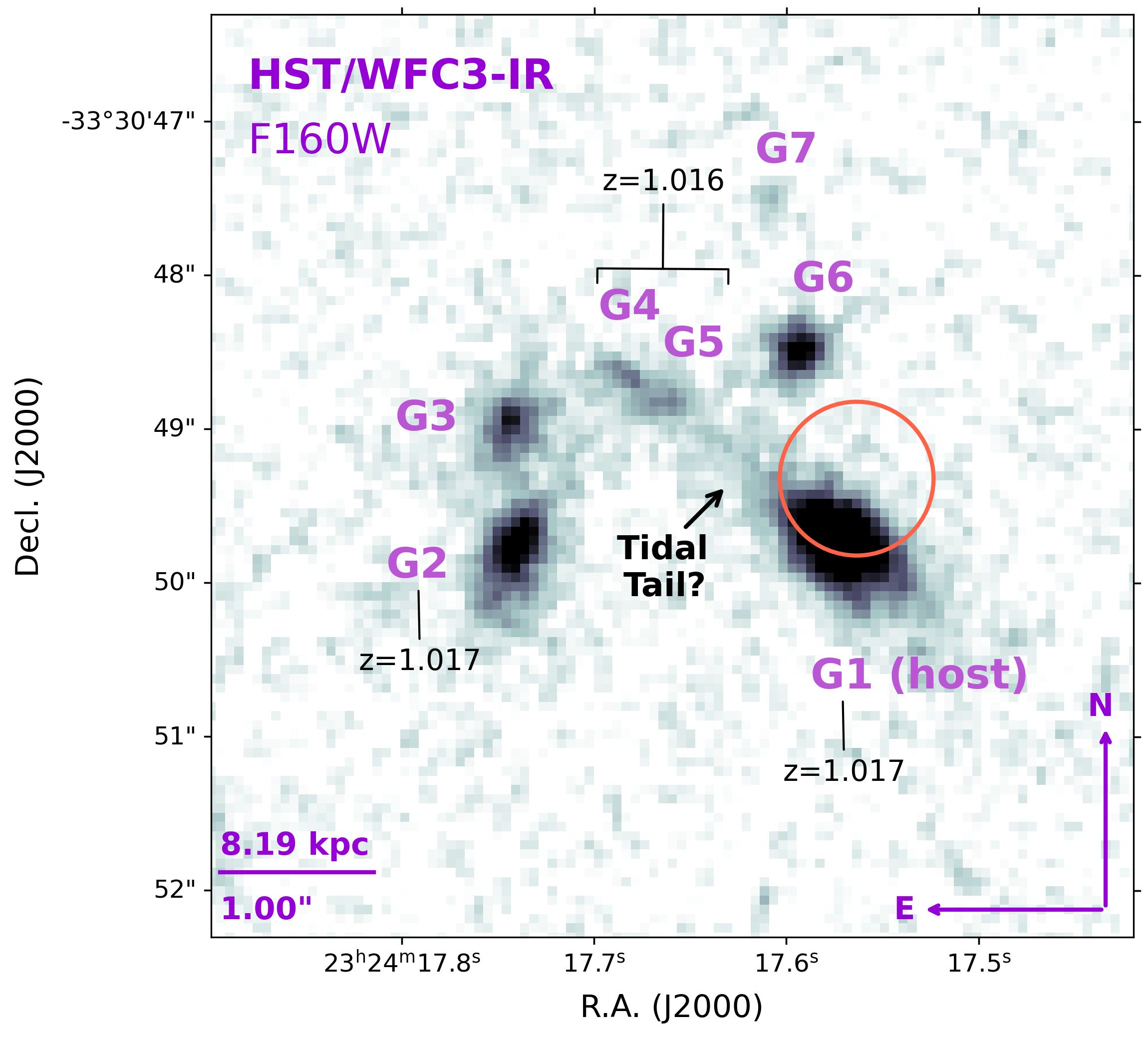}
    \caption{\textit{HST}/WFC3-UVIS F606W (left) and WFC3-IR F160W (right) imaging of the host galaxy of FRB\,20220610A. The sensitivity and depth afforded by \textit{HST} reveals multiple galaxies, labeled clockwise G1-G7, where G1 is the most likely host galaxy. The FRB localization (orange ellipse, 68\% confidence) is coincident with G1. We annotate the image with the spectroscopic redshifts of G1, G2, G4, and G5 from the X-Shooter observations (first published in \citet{Ryder23}, described in Section~\ref{sec:system} and Appendix~\ref{app:emission-lines}) and denote the location of potential tidal tails between G1 and G5 evident in both filters. We smooth the F606W image to achieve comparable resolution to the F160W image, which is unsmoothed.}
    \label{fig:imaging}
\end{figure*}

\citet{Ryder23} identified a ``complex host galaxy system'' in their ground-based imaging of FRB\,20220610A. They identified it as either one galaxy with three distinct clumps or 2--3 separate galaxies potentially interacting or even merging (their Figure~2). With the spatial resolution and sensitivity afforded by \textit{HST}, we find that the complex identified in their imaging is in fact composed of at least five extended sources, with two additional sources detected in the vicinity that were not previously visible. We label these sources numerically as 1-7, where G1 is the source coincident with the FRB ($68\%$ confidence) and is clearly detected in both \textit{HST} filters. G2-G7 follow clockwise from G1 (Figure~\ref{fig:imaging}). We note that all sources are clearly extended with respect to the instrumental PSF, indicating that they are galaxies, except for G7, which is impossible to distinguish from a point-like source. We note that G3 is not detected in the F606W imaging, signifying that it is comparatively redder. We further note the presence of faint extended emission between G1 and G5 in both filters, possibly indicating a stream or sign of tidal interaction.

We refine the analysis of \citet{Ryder23} to assess the true host galaxy of FRB\,20220610A based on this {\it HST} imaging. They determined a 99.99\% probability of association with ``component (a)'' (roughly, G1 here). We use the Bayesian framework PATH (Probabilistic Association of Transients to their Hosts; \citealt{path}) to determine the likelihood of association between the FRB and each of the seven galaxies identified in the F606W imaging. PATH uses the FRB localization, the galaxy's sky position, half-light radius, apparent magnitude, and the FRB's galactocentric offset and probability that the host is undetected in the image, $P(U)$, which we assume to be 0 given the depth of the \textit{HST} imaging. PATH then provides the posterior of the probability of association to the input list of galaxies, $P(O\vert x)$, for which higher values correspond to a higher certainty of association.

We use the F606W image for this analysis as it is most proximate in wavelength to observed $r$-band, on which PATH priors are calibrated. G1, the closest galaxy to the FRB, has $P(O\vert x) = 0.87$. G6 is the next closest in proximity and has $P(O\vert x) = 0.11$. All remaining galaxies have $P(O\vert x) < 0.016$. In testing larger values of $P(U)$ (from 0.1 to 0.9), we find a negligible effect on the posteriors; thus, our assumption of $P(U)=0$ is well-justified. We thus conclude that G1 is the most likely host of FRB\,20220610A and present the $P(O\vert x)$ for each of the seven galaxies in Table~\ref{tab:host/gal_props}. We note in the less probable case (by a factor of $\approx8$) that G6 is the true host, the photometric redshift of G6 is consistent within uncertainties with the spectroscopic redshift of G1 (see Section~\ref{sec:SPP-modeling}). Moreover, follow-up spectroscopy of G6 indicates a redshift consistent with $z \approx 1$ (see below). With \texttt{Source Extractor}, we derive a position for G1 of R.A.=\ra{23}{24}{17.573}, decl.=\dec{-33}{30}{49.725} (J2000) in the F160W filter.

Finally, to determine the spectroscopic redshifts of the galaxies in the system, we revisit the spectra taken by \citet{Ryder23} with VLT/X-Shooter. These data were taken at two position angles of 45$^{\circ}$ and 90$^{\circ}$ East of North and totaled 7200 and 7548~s of exposure, respectively. When overlaid on the F160W image, the two slit orientations cover the locations of G1, G2, G4, and G5 (Figure~\ref{fig:emission-line-fig}). Using the {\it HST}-determined positions, we identify the [O\,II]$\lambda$3726\AA, 3729\AA~doublet at the expected locations of the galaxies. Unfortunately, the expected location of the H$\alpha$ emission line falls on a telluric sky line near 1.3~$\mu$m; while H$\alpha$ is apparent in the 2D spectra (giving enough confidence for a redshift; c.f. \citealt{Ryder23}), we cannot reliably estimate its flux. However, based on the [O\,II] doublet, we establish the redshifts of G1 and G2 at $z=1.017$. As the contributions of G4 and G5 are blended in the X-Shooter spectrum (they are not even clearly separated in our {\it HST} imaging), we determine the redshift of the blended galaxies at $z=1.016$. Finally, we obtained a third X-Shooter spectrum on 15 November 2023 UTC (PI Shannon, Program 108.21ZF.012) covering the locations of G1, G6, and G7; we find probable [O\,II] emission from G1 and G6 consistent with $z \approx 1$. Further details on the spectra and emission line identification for these galaxies are presented in Appendix~\ref{app:emission-lines}. 

The lack of detection of G3 in F606W, coupled with the detection in F160W, is intriguing in the context of a high-redshift origin. Indeed, if we interpret the red color to be attributed to the 4000\,\AA\ break, this results in a redshift range of $z\approx 0.5-2.7$. However, with present data we cannot rule out the possibility that the reddening is due to a combination of dust extinction or an older stellar population at a different redshift.

\subsection{PSF Modeling and Galfit Analysis} \label{sec:psf_and_galfit}

To quantify the sizes and morphologies of the host and surrounding galaxies in the \textit{HST} images, as well as to determine the presence of any sub-structure, we use the galaxy profile modeling software \textsc{Galfit} (v3.0.5) \citep{galfit_2002,galfit_2010}. We derive the empirical PSF of the {\it HST} images (a required input for \textsc{Galfit}) with the \texttt{hst\char`_wfc3\char`_psf\char`_modeling} software package \citep{PSF_zenodo,Revalski_etal_2023}, employing the \texttt{stellar} option to allow a user-input star catalog. After manually inspecting the images to select stars with clear Airy disk structure for the PSF model, we use \texttt{Source Extractor} to derive the star positions and input this catalog to the PSF modeling code. For the F606W image, we use five stars, resulting in a FWHM = 3.033 pixel PSF ($0\farcs091$); for the F160W image, we use six stars, deriving a PSF FWHM of 3.699 pixels ($0\farcs222$). We use the median PSF model (as opposed to the mean model) for all further analysis as this minimizes the degree of contribution from neighboring sources in the fields of the individual PSF stars.

\textsc{Galfit} takes initial values on the position, size, magnitude, semi-major/-minor axis ratio, and position angle of a galaxy. It then convolves the data with the model (incorporating the empirical PSF) to return the best-fit values for the input parameters. We model the galaxies with a S\'ersic surface brightness profile \citep{Sersic_1968}
\begin{equation}
    I(r) = I_e \times {\rm exp}(-b_n[(r/r_e)^{1/n} - 1]),
\end{equation}
where $r_e$ is the effective radius that encompasses half of the galaxy's light, $I_e$ is the surface brightness at $r_e$, $n$ is the S\'ersic index which describes the shape of the light profile (e.g., $n=1$ for an exponential disk, $n=4$ for an elliptical de Vaucouleurs profile), and $b_n$ is a variable coupled to $n$ that ensures $r_e$ encompasses half of the galaxy's flux. While we fit G1 with a single S\'ersic component in F606W, we use two S\'ersic profiles in F160W (nominally representing bulge and disk components) as this produces cleaner residuals. We initialize the bulge and disk profiles with different effective radii and S\'ersic indices to reflect the expected size and light profiles of the two components. We then fit the remaining galaxies in each filter with single S\'ersic profiles, with the exception of G2 in the F606W image which is similarly modeled as two components to improve the residual.

\begin{figure*}
    \centering   
    \includegraphics[width=\textwidth]{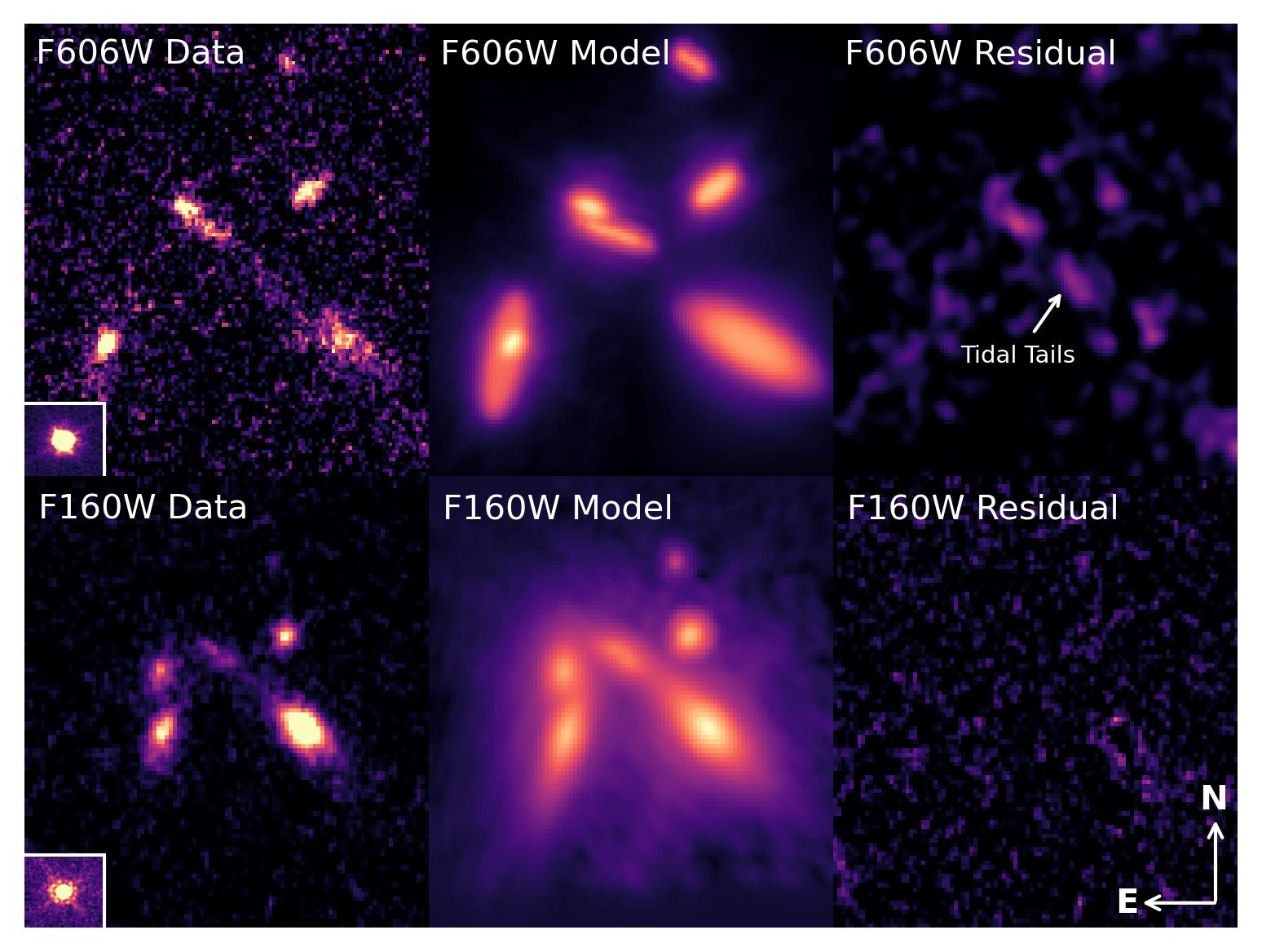}
    \caption{\textsc{Galfit} modeling of the FRB\,20220610A host and surrounding galaxies in F606W (top row) and F160W (bottom row). For each filter, we show the original science image (unsmoothed, left), the \textsc{Galfit} model (center), and the residual (right). The PSF derived for each filter is shown in the lower left corner of the science images. In F606W, we fit G1 and G4-G7 with single S\'ersic profiles and model G2 with two S\'ersic profiles, representing bulge and disk components. In F160W, we fit G1 with two S\'ersic profiles (bulge and disk); the remaining galaxies are modeled with single S\'ersic profiles. While possible tidal interaction is visible between G1 and G5 in both of the science images (left column), this feature is only apparent in the F606W residual. We smooth the F606W residual to highlight this structure.}
    \label{fig:galfit}
\end{figure*}

We present the results from the \textsc{Galfit} analysis, including the values for $n$ and $r_e$, in Table~\ref{tab:host/gal_props}. The original science images, \textsc{Galfit} models, and residuals are shown in Figure~\ref{fig:galfit}. While the science images in both filters show extended emission between G1 and G5 along the northeast-southwest direction, referred to here as tidal tails, this structure is only apparent in the F606W residual. We comment on the implications of these features in Section~\ref{sec:disc-gal-comp}.

\begin{deluxetable*}{ccccccccccc}
\tabletypesize{\footnotesize}
\tablewidth{0pc}
\tablecaption{FRB\,20220610A Host and Surrounding Galaxy Properties
\label{tab:host/gal_props}}
\tablehead{
\colhead{Source} &
\colhead{R.A.} &
\colhead{Decl.} &
\colhead{Filter} &
\colhead{S\'ersic Index} &
\colhead{$r_{\rm e}$} &
\colhead{$z$} &
\colhead{PATH} & 
\colhead{log(M$_*$/M$_{\odot}$)} &
\colhead{$t_{\rm m}$} &
\colhead{${\rm SFR}_{\rm 0-100~Myr}$} \\ 
\colhead{} &
\colhead{(J2000)} &
\colhead{(J2000)} &
\colhead{} &
\colhead{$n$} &
\colhead{\arcsec} &
\colhead{} &
\colhead{} &
\colhead{} &
\colhead{(Gyr)} &
\colhead{(M$_{\odot}$~yr$^{-1}$)}
}
\startdata
1 & 23:24:17.573 & $-$33:30:49.725 & F606W & 0.32 & 0.33 & 1.017$^{a}$ & 0.87 & 9.69$\pm$0.11 & $2.60^{+0.61}_{-0.91}$ & $1.67^{+2.41}_{-0.95}$ \\
 & & & F160W (disk) & 0.55 & 0.41 & & & & \\
 & & & F160W (bulge) & 1.67 & 0.26 & & & & \\
2 & 23:24:17.743 & $-$33:30:49.796 & F606W (disk) & 0.07 & 0.33 & 1.017$^{a}$ & 6.9e-4 & 9.19$^{b}$ & & \\
 & & & F606W (bulge) & 0.05 & 0.06 & & & & \\
 & & & F160W & 1.59 & 0.34 & & & & \\
3 & 23:24:17.744 & $-$33:30:48.982 & F160W & 2.25 & 0.25 & & 5.3e-5 & 8.40$^{b}$ & & \\
4 & 23:24:17.683 & $-$33:30:48.657 & F606W & 0.39 & 0.08 & 1.016$^{a,c}$ & 4.9e-3 & 7.83$^{b}$ & & \\
 & & & F160W & 1.23 & 0.20 & & & & \\
5 & 23:24:17.665 & $-$33:30:48.813 & F606W & 0.09 & 0.22 & 1.016$^{a,c}$ & 1.6e-2 & 8.58$^{b}$ & & \\
 & & & F160W & 1.59 & 0.27 & & & & \\
6 & 23:24:17.596 & $-$33:30:48.485 & F606W & 0.04 & 0.14 & $0.72^{+0.92}_{-0.39}$$^{d}$ & 0.11 & $8.71^{+0.58}_{-0.57}$ & $5.56^{+1.51}_{-1.94}$ & $0.10^{+1.10}_{-0.10}$ \\
 & & & F160W & 0.44 & 0.08 & & & & \\
7 & 23:24:17.609 & $-$33:30:47.506 & F606W & 0.06 & 0.08 &  & 1.6e-4 & 6.10$^{b}$ &  &  \\
 & & & F160W & 0.01 & 0.10 & & & & \\
\enddata
\tablecomments{The properties of the host of FRB\,20220610A and surrounding galaxies in the F606W and F160W images, with coordinates determined from F160W. The S\'ersic index $n$ and effective radius $r_{\rm e}$ are derived from \textsc{Galfit}. The redshifts were determined via X-Shooter spectroscopy or photo-z modeling with \texttt{Prospector}. The PATH column denotes the probability posterior $P(O\vert x)$ for association with the FRB. We list the stellar mass (log(M$_*$/M$_{\odot}$)), mass-weighted age ($t_{\rm m}$), and recent SFR integrated over the past 100~Myr in lookback time (SFR$_{\rm 0-100~Myr}$) from \texttt{Prospector}. \\
$^{a}$ Spectroscopic redshift \\
$^{b}$ Derived via \citet{van_der_Wel+14} mass-radius relation (see Section~\ref{sec:disc-DMexcess}). \\
$^{c}$ Galaxies 4 and 5 are blended and unable to resolve separately, so we report the redshift of the blended object \\
$^{d}$ Photometric redshift derived via SED modeling \\
}
\end{deluxetable*}

\subsection{Aperture Photometry} \label{sec:photom}

Based on our {\it HST} imaging where each galaxy near the localization region of FRB\,20220610A is resolved (Figure~\ref{fig:imaging}), we note that their apparent sizes are 0.08--0.3\arcsec, smaller than the size of the PSF in our ground-based VLT and Keck imaging. Therefore, to estimate accurate fluxes for each separate galaxy using an aperture smaller than the PSF, we need to implement aperture corrections within each ground-based image. This method is ideal for estimating accurate fluxes of unresolved sources because it allows us to: (1) infer an accurate flux of each source to within the measurement uncertainties of our photometry and aperture correction, and (2) deblend emission from galaxies where their separations are close to the PSF size within each image. Below we describe a procedure for estimating aperture corrections using a curve-of-growth method within each ground-based image as well as estimating the flux from each galaxy.

We first astrometrically tie the {\it HST} imaging to the VLT $R_{\rm special}$-band imaging as described in Section~\ref{sec:hst-obs}. We then derive the positions of the galaxies using {\tt Source Extractor} from the F160W image, where all galaxies are detected and resolved (listed in Table~\ref{tab:host/gal_props}). Next, we verify the astrometric alignment of the ground-based images by overlaying the galaxy positions. In the case of the $J$-band image, the positions were slightly misaligned, so we used a star in the field to calculate the offset required to shift the coordinates to the correct location. We then perform photometry on our \textit{HST} imaging and the imaging from \citet{Ryder23} (c.f., Section~\ref{sec:obs}) at the {\it HST} positions of the galaxies using a custom script that implements the \texttt{aperture\char`_photometry} module of \texttt{photutils} \citep{photutils}\footnote{https://github.com/charliekilpatrick/photometry}. We use larger aperture sizes for the space-based images compared to the ground-based images to reduce the degree of blending. 

Next, we estimate the curve-of-growth using point sources within each ground-based image and use this to apply an aperture correction to obtain accurate flux measurements. The curve-of-growth represents the fractional encircled flux $f$ (i.e., the encircled energy correction) of a typical point source within the image as a function of the aperture radius (i.e., $r$ in arcsec) used.  Thus for an aperture radius of a given value, we can estimate the total flux from an unresolved source by applying an aperture correction $2.5\log_{10}f$.  We use several point sources to estimate the curve-of-growth, normalizing by the total flux for that source out to large radii; this is typically $r>$3~times the FWHM, where we would expect to encompass $>$99.7\% of the light. In this way, we estimate the average value and standard deviation for the curve-of-growth at each value of $r$, which we then transform into the aperture correction and uncertainty in the aperture correction at a given aperture radius. We emphasize that, in addition to the measurement uncertainties on the aperture correction itself, the Poisson uncertainties from the aperture photometry below are also increased by the inverse of the encircled energy correction $1/f$, and propagated.

Finally, we apply our aperture photometry procedure and curve-of-growth measurement to G1 and G6 for use in modeling their stellar population properties (see Section~\ref{sec:SPP-modeling}). For G1, we use a 1$\arcsec$ aperture for the \textit{HST} imaging and a $0\farcs5$ aperture with curve-of-growth corrections for the ground-based imaging. Similarly for G6, we use a larger $0\farcs4$ aperture for the \textit{HST} images and a smaller $0\farcs3$ aperture for the ground-based images. We select a source-free region directly to the west of the host system to estimate the background level via an annulus at twice and eight times the size of the aperture to avoid contamination from the surrounding galaxies in the FRB\,20220610A system. In cases in which the galaxies are not detected in the imaging, we derive $3\sigma$ upper limits by performing photometry on faint sources in the field. We then correct the photometric values and limits for Galactic extinction using the \citet{Fitzpatrick:2007} extinction law with $E(B-V)=0.0158$ derived via the \citet{SandF} dust map. All photometric measurements are presented in Table~\ref{tab:Obs-log}. We find detections at $\gtrsim 3\sigma$ significance for G1 in the $g$, $V$, F606W, $R_{\rm special}$, $J$, F160W, and $K_{\rm s}$ filters; and for G6 in the F606W, $R_{\rm special}$, $J$, and F160W filters. All galaxies have $\gtrsim 3\sigma$ detections in the \textit{HST} images except G3 in F606W where it is not detected. We derive upper limits in each of the remaining photometric bands.

\subsection{Stellar Population Modeling} \label{sec:SPP-modeling}

To derive the stellar population properties of G1 and G6, we use the Bayesian inference code \texttt{Prospector} \citep{Johnson+21}. We initialize the model using the same prior assumptions as in \citet{Gordon+23} to fit the nine available bands of photometry (seven detections, two upper limits) of G1 as presented in Table~\ref{tab:Obs-log}. We fit all available photometry; the lack of continuum in the VLT spectra (see Appendix Figure~\ref{fig:emission-line-fig} and \citealt{Ryder23}) prohibits inclusion in the fits. The best-fit model for G1 has a stellar mass of log(M$_*$/M$_{\odot}$) $=$ 9.69$\pm$0.11, a mass-weighted age of $t_{\rm m}$ $=2.60^{+0.61}_{-0.91}$~Gyr, and a star formation rate integrated over the past 100~Myr of ${\rm SFR}_{\rm 0-100~Myr}$ $=1.67^{+2.41}_{-0.95}$~M$_{\odot}$~yr$^{-1}$ (68\% uncertainties). We derive a specific star formation rate (star formation rate per unit stellar mass) of log(${\rm sSFR}_{\rm 0-100~Myr}$)=$-9.41^{+0.44}_{-0.41}$~yr$^{-1}$. Per the mass-doubling number criterion from \citet{Tacchella+22}, 
this galaxy is classified as star-forming 
(based on the specific star formation rate and the redshift; applied in previous work on FRB hosts; e.g., \citealt{Gordon+23,Sharma_etal_2023,Ibik+23}). 
We report the median and 68\% confidence intervals of the stellar population properties in Table~\ref{tab:host/gal_props}.

Of the seven sources identified in the {\it HST} imaging, the VLT spectroscopy from \citet{Ryder23} covers all except G3, G6, and G7. While G3 is possibly at $z \gtrsim 0.5$ to explain its red color (see Section~\ref{sec:disc-gal-comp}), the lack of additional detections beyond a single band precludes further inference. While G7 is detected in both filters, there is not enough data to reliably model this source. However, for G6, the combination of {\it HST} data and detections recovered from ground-based imaging results in four photometric measurements (of the nine bands), enabling a \texttt{Prospector} fit to estimate its redshift (and additionally, derive its stellar population properties). This is particularly relevant as it is the next most likely host of FRB\,20220610A ($P(O\vert x) = 0.11$). We initialize the G6 model using the four detections and five upper limits with redshift as a free, uniform prior between $z=0$ and $z=3$. The best-fit model suggests a galaxy with redshift $z=0.72^{+0.92}_{-0.39}$ and log(M$_*$/M$_{\odot}$) $=8.71^{+0.58}_{-0.57}$. We report these stellar population properties and additional properties in Table~\ref{tab:host/gal_props}. This is fully consistent with the redshifts of the other sources, albeit with large uncertainties (but see Appendix~\ref{app:emission-lines} for a probable $z\approx 1$ origin).

\subsection{The Location of FRB\,20220610A within its Host} \label{sec:frac-flux}

We first measure the physical offset of FRB\,20220610A with respect to the center of G1 as reported in Table~\ref{tab:host/gal_props}. We find the angular separation between the FRB location and the host optical center (via the F606W image) is $0\farcs36 \pm 0\farcs71$, corresponding to a projected physical offset of 2.97$\pm$5.86 kpc. The uncertainty in this measurement encompasses the uncertainties in the host position, absolute astrometry, and FRB position\footnote{We note that the ``negative'' offset as allowed by the uncertainty can be physically interpreted as the FRB originating to the south-east of the host nucleus.}. We also estimate the angular offset in a different way following the procedure in \cite{Mannings+21} which incorporates the size of the localization region via a weighting scheme. First, we calculate a distribution of offsets within a 3$\sigma$ localization region centered at the FRB position, where $\sigma$ is the uncertainty on the FRB localization. Next, by weighting each offset with a 2D Gaussian probability distribution, we derive an average angular offset of $0\farcs70 \pm 0\farcs36$, in which the uncertainty is represented by the standard deviation of the distribution; this is equivalent to a physical offset of 5.78$\pm$2.97 kpc. The angular offset derived from the Gaussian weighting is likely inflated by the large FRB localization uncertainties which introduce a spread in the distribution. This spread increases the influence of large offsets, thereby shifting the average away from the central value of the distribution. However, this weighted offset more aptly reflects the true value given the size of the FRB\,20220610A localization region.

To quantify the location of FRB\,20220610A with respect to its host's light distribution, we next estimate the fractional flux (F$_{\rm F}$; e.g., \citealt{Mannings+21}) in the \textit{HST} images. Fractional flux values range from 0--1, with a value of $F_{\rm F}=1$ indicating the FRB is at the brightest position within the galaxy with respect to the light distribution. F606W probes rest-frame UV light at the redshift of FRB\,20220610A, and can be interpreted as tracing recent star formation via hot, young stars. F160W probes the rest-frame optical light, which physically maps to older stellar populations, and is thus a proxy for the stellar mass distribution of the galaxy. Using the methodology developed in \citet{Mannings+21}, we find the fractional flux at the position of FRB\,20220610A to be 0.35$\pm$0.30 in F606W and 0.38$\pm$0.28 in F160W. These values are consistent within the 68\% confidence intervals of rest-frame UV and optical fractional flux for the 10 FRB hosts with {\it HST} imaging analyzed by \citet{Mannings+21}. The large uncertainties, owed mainly to the size of the localization region relative to the host size, preclude a robust indication that the FRB position traces the star formation and stellar mass distributions of its host galaxy. However, it is likely that FRB\,20220610A resides in an average or below average light location compared to the total distributions in star formation and stellar mass.

\section{Discussion} \label{sec:disc}
\subsection{The Host of FRB\,20220610A as a Member of a Compact Group} \label{sec:disc-system-ID}

We first explore the large-scale environment of FRB\,20220610A as identified in our \textit{HST} imaging. The host galaxy, G1, is proximate to six additional sources within a radius of $\approx$16~kpc about the center of the complex, the majority of which are spectroscopically confirmed to be at near-identical redshifts. The proximity of the galaxies to each other and their comparable redshifts suggests membership in a larger system. Thus, to put the large-scale environment of FRB\,20220610A in the context of similar galaxies at this redshift, we first attempt to categorize the system. 

On the largest and most massive scale we consider galaxy clusters. These systems encompass total masses (defined as the combined stellar and dark matter halo mass) of $\gtrsim10^{14}$ M$_{\odot}$ and span a few Mpc on the sky \citep{Bahcall99}. At $z \approx 1$, a galaxy cluster would have an angular size of $\gtrsim$2$\arcmin$, and thus it is not possible to visually identify an over-density given the FOV of our \textit{HST} imaging. Therefore we use known galaxy cluster surveys to determine whether the collection of galaxies is part of a previously-identified, larger-scale cluster. Galaxy clusters are typically identified via three methods \citep{Oguri+18}: (1) optical identifications of galaxy overdensities with wide-field surveys; (2) X-ray detections of hot gas between galaxies; and/or (3) radio/millimeter observations of the thermal Sunyaev-Zel'dovich (SZ; \citealt{SZ_effect}) effect in which electrons between the galaxies scatter cosmic microwave background photons via inverse Compton scattering. Thus, detections of the SZ effect can be used to infer the presence of galaxy clusters through the hot gas in the intracluster medium. 

However, public optical surveys are limited in either their redshift reach ($z\lesssim0.9-1.1$; \citealt{Oguri+18}) or spatial coverage. Indeed, one of the most promising surveys uses the 8.2m~Subaru telescope's Hyper Suprime-Cam, but unfortunately it does not cover the southern location of FRB\,20220610A \citep{Oguri+18}. We also examine X-ray sources in the ROSAT All-Sky-Survey (2RXS) within the ROSAT All-Sky-Survey Multi-Component Matched Filter catalog of X-ray selected galaxy clusters \citep{2RXS,Klein+23}. This catalog has a purity of 90\%, meaning 90\% of the sources identified in the catalog are indeed galaxy clusters to high probability. While the catalog extends to $z\approx1$ and covers the location of FRB\,20220610A, there is no identified source in 2RXS within $\approx 1^{\circ}$ of the FRB\,20220610A system. However, as the majority of sources identified in the catalog are at $z\lesssim0.4$, the completeness of the survey is poor at the redshift of FRB\,20220610A.

We conclude our cluster search with available SZ-selected galaxy cluster surveys. Of these, the Planck 2nd Sunyaev-Zeldovich Source Catalog \citep{Planck_SZ} and a SZ-selected cluster survey based on the Atacama Cosmology Telescope 5th Data Release (ACT DR5; \citealt{ACT_DR5_SZ}) have both the sky and redshift coverage to identify a cluster associated with FRB\,20220610A. The Planck catalog (which has a purity of 83-87\%) does not identify a cluster within 20$\arcmin$ of FRB\,20220610A, a conservative upper limit on the size of a galaxy cluster at this redshift. This survey has an 80\% total mass completeness limit of $\approx7.5\times10^{14}$ M$_{\rm 500}/{\rm M}_{\odot}$ at $z\approx1$, a common mass estimate in the cluster literature representing the mass within a spherical radius encompassing 500 times the critical density at the cluster's redshift. Similarly, no cluster is identified in ACT DR5 within 20$\arcmin$. This survey has a 90\% total mass completeness limit of 3.2$\times10^{14}$ M$_{\rm 500c}/{\rm M}_{\odot}$ at $z\approx1$ (where M$_{\rm 500c}$ is another way to denote the critical mass density).

\begin{figure*}
    \centering
    \includegraphics[width=0.7\textwidth]{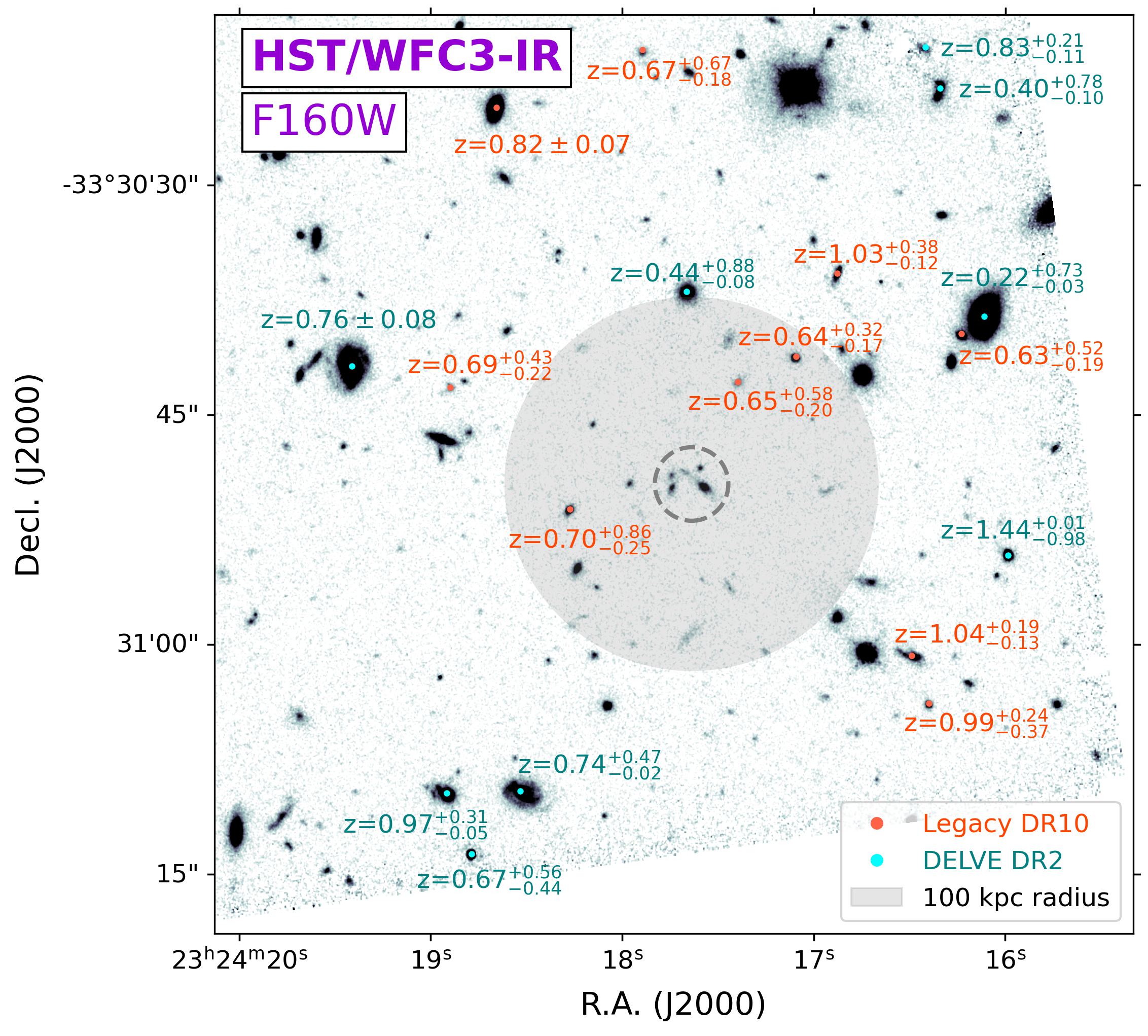}
    \caption{$60\arcsec \times 60\arcsec$ field of view centered on the FRB\,20220610A compact group (dashed circle) with a 100~kpc transparent grey region about the group for scale. All available photometric redshifts (median $\pm$ 68\% confidence), when filtered for galaxies, from DELVE~DR2 are plotted in green and supplemented with Legacy Survey DR10 redshifts (orange). While individual redshifts exhibit large uncertainties, the majority of photometric redshifts are at $z \lesssim 1$. This indicates that FRB\,20220610A is not likely part of a larger galaxy cluster, and is instead in a standalone, compact group.}
    \label{fig:photo-zs}
\end{figure*}

To determine if a galaxy cluster \textit{could} be identified, we obtained photometric redshifts of objects in the vicinity of FRB\,20220610A from the DECam Local Volume Exploration Survey DR2 (DELVE; \citealt{DELVE_DR2}) and the Legacy Survey DR10 \citep{Legacy}. We apply a star-galaxy classifier on the Legacy data based on the PSF of the object and its classification in the Gaia survey to select only on high-probability galaxies. In Figure~\ref{fig:photo-zs}, we show the surrounding $\approx$60\arcsec\ of the FRB\,20220610A system (corresponding to 164~kpc at $z=1.017$) with photometric redshifts overlaid. While this region is not large enough to observe the full extent of a large cluster at $z\approx1$ (as we are limited by the field of view of the \textit{HST} imaging), any apparent over-density in redshifts would be noticeable on this scale. The DELVE redshifts have higher precision and all have median values of $z \lesssim 1$. The same is true for the supplemented Legacy catalog redshifts although their redshift uncertainties are much larger. Overall, we find no apparent large over-density of galaxies at $z\approx 1$, and thus we conclude there is no obvious sign of a galaxy cluster at this location. 

At smaller masses than galaxy clusters are galaxy groups. As the FRB\,20220610A system is comprised of several galaxies at comparable redshifts within tens of kiloparsecs, its properties are instead consistent with those of compact groups (CG; \citealt{Hickson1982,Hickson1997,Sohn_etal_2016}). While CGs have mainly been studied at low redshift owing to the difficulty in identifying them (i.e., \citealt{Hickson1982,Coenda_etal_2012, zheng_and_shen20}), galaxy simulations have shown that the fraction of galaxies existing in CGs peaks at $z\approx1.0-2$ \citep{wiens_etal_2019}. However, these systems are still rare compared with typical field galaxies, with only 0.1$-$1\% of galaxies residing in compact groups at these redshifts \citep{dg20}. We further comment on the implications of the host of FRB\,20220610A belonging to such an environment in Section~\ref{sec:disc-cggal-comp}.

\subsection{The Host in the Context of Field Galaxies} \label{sec:disc-gal-comp}

\begin{figure}
    \centering   
    \includegraphics[width=0.5\textwidth]{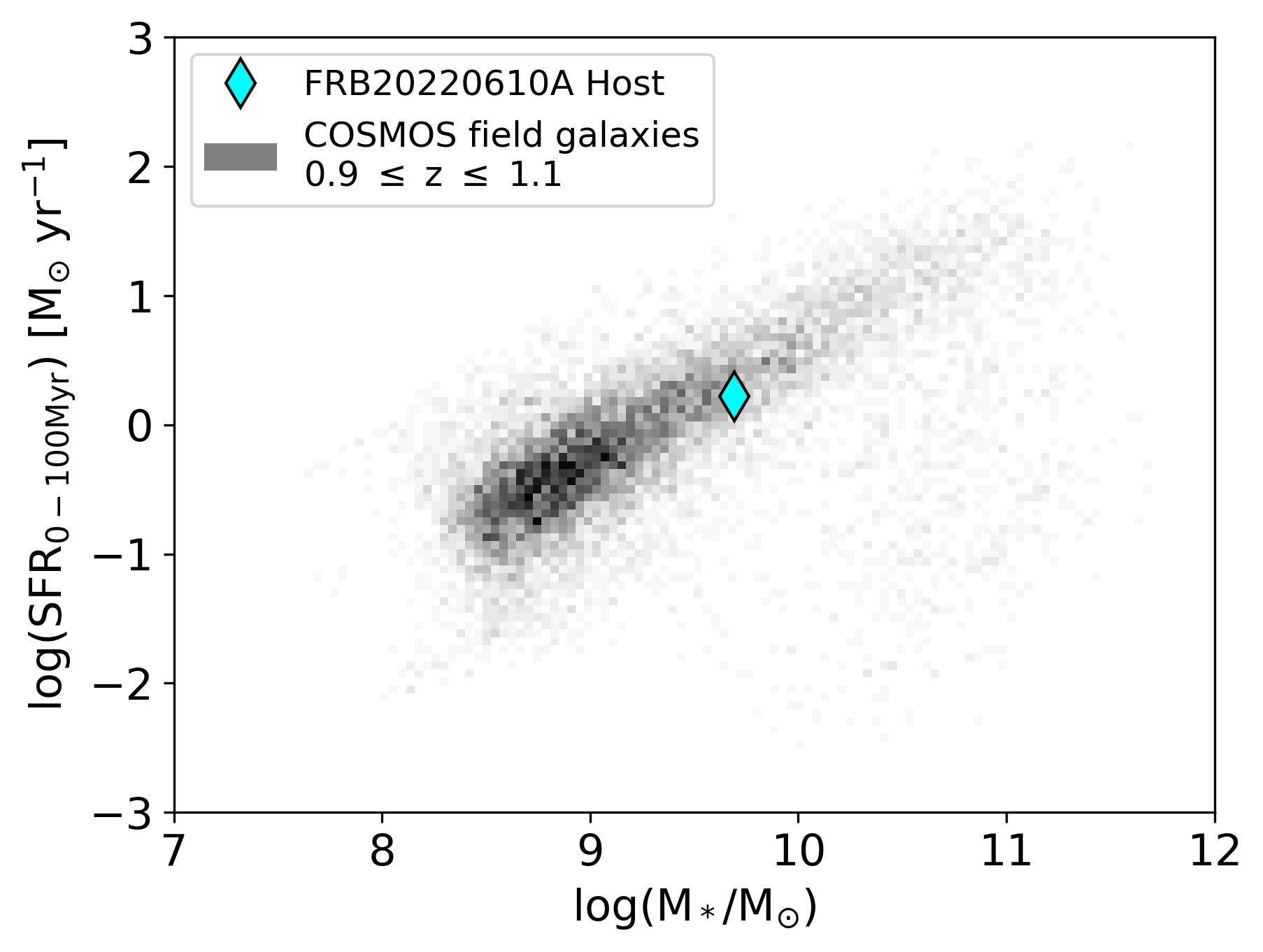}
    \caption{The star formation rate (integrated over the last 0--100~Myr) versus stellar mass parameter space denoting the star-forming main sequence (SFMS). The host of FRB\,20220610A (cyan diamond) is plotted against field galaxies from the COSMOS survey spanning $0.9\leq z \leq 1.1$ \citep{Laigle_etal_2016,Leja_etal_2020}. The host galaxy traces the SFMS, indicating it is actively star forming.}
    \label{fig:SFMS}
\end{figure}

We next contextualize the galactic and large-scale environment of FRB\,20220610A. The lack of FRB hosts in the regime of $z\approx 1$ prevents an accurate comparison to that of FRB\,20220610A, as galaxy properties evolve with redshift. Instead, we leverage deep galaxy surveys to compare the host of FRB\,20220610A to field galaxies at comparable redshifts to assess whether it traces galaxy properties in the same way as the low-$z$ FRB host population. We compare to field galaxies at $z\sim 1$ from the COSMOS sample \citep{Laigle_etal_2016} whose stellar population properties were derived with \texttt{Prospector} with similar prior assumptions to our work, thus enabling a direct comparison mainly free of systematic differences due to modeling \citep{Leja_etal_2020}. We construct a distribution of COSMOS galaxies with redshifts $\pm$0.05 about the host redshift of $z=1.017$, resulting in a sample of 3976 field galaxies. We find the stellar population properties of this sample have medians and 68\% confidence intervals of log(M$_*$/M$_{\odot})=9.14^{+0.96}_{-0.47}$, $t_{\rm m}=2.51^{+0.52}_{-0.53}$, and SFR$_{\rm 0-100~Myr}=0.73^{+2.86}_{-0.54}$. Within the uncertainties, the stellar population properties of the host of FRB\,20220610A is consistent with the field galaxy population. 

In Figure~\ref{fig:SFMS}, we consider the host's placement relative to these COSMOS field galaxies at similar redshifts on the star-forming main sequence (SFMS) in the stellar mass--SFR parameter space.  Previous studies \citep{Gordon+23,Law+23,Ibik+23} have performed similar analyses by comparing the location of FRB host galaxies relative to the SFMS against a background sample of field galaxies from the COSMOS sample.  These works found that most FRB hosts at $z\lesssim0.5$ trace the SFMS. Similar to the low-$z$ FRB hosts, we find that the host of FRB\,20220610A falls on the SFMS (Figure~\ref{fig:SFMS}), implying that it is forming stars at a similar rate to galaxies of comparable stellar mass at its redshift. We note, however, that it is slightly more massive than the majority of star-forming galaxies in this redshift range. In the absence of a conclusive star formation history (SFH) derived for the host (as the signal-to-noise of the data results in large uncertainties in the star formation rate in each age bin), we indirectly compare the SFH to the sample of field galaxies using the log(${\rm sSFR}_{\rm 0-100~Myr}$)--$t_{\rm m}$ parameter space (c.f., \citealt{Gordon+23}). These parameters are moments of the SFH and serve as a proxy to a direct SFH comparison. We find the host of FRB\,20220610A traces the COSMOS field galaxies in this parameter space as well. 

\begin{figure}[t]
    \centering   
    \includegraphics[width=0.5\textwidth]{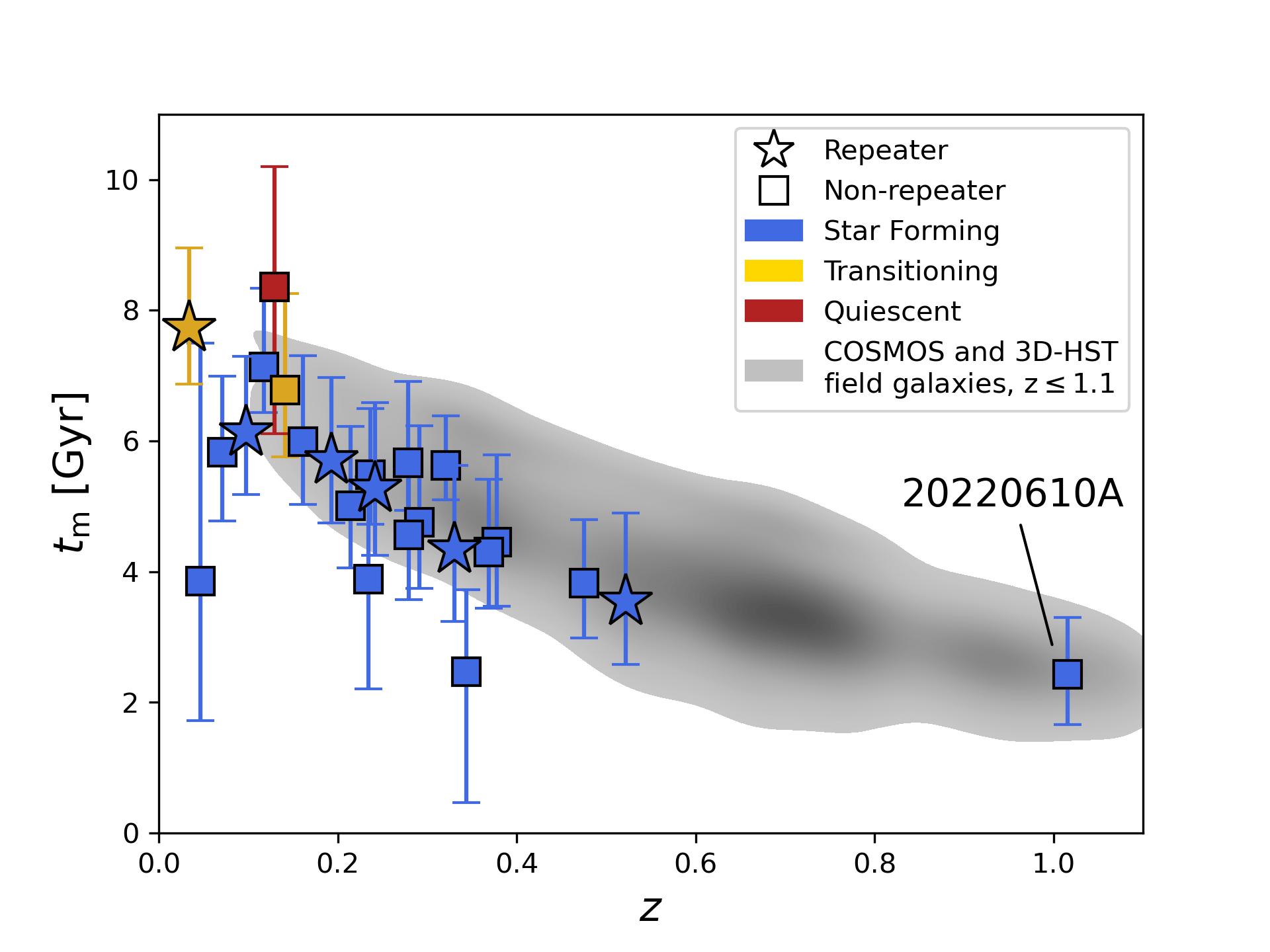}
    \caption{Mass-weighted age versus redshift for all FRB hosts with reported mass-weighted ages \citep{Gordon+23,Ibik+23} against COSMOS and 3D-HST \citep{3D_HST,Leja_etal_2020} field galaxies. FRB hosts trace the age of typical galaxies in the Universe across redshift, and the host of FRB\,20220610A follows this trend. We represent repeaters with stars and non-repeaters with squares and color the FRB hosts by their degree of star formation per \citet{Tacchella+22}'s mass doubling number classification (e.g., their Eqn.~2). The overdensity in the COSMOS field galaxies at $z\approx0.7$ is a product of the redshift range shown.}
    \label{fig:age-vs-z}
\end{figure}

Finally, we compare the mass-weighted age ($t_{\rm m}$) versus redshift of the host of FRB\,20220610A with other FRB hosts from the literature against COSMOS field galaxies in Figure~\ref{fig:age-vs-z} (data from \citealt{Gordon+23,Ibik+23}). As redshift increases, the evolution from large to small absolute ages is a reflection of the galaxy population becoming younger at higher redshifts; in other words, if the mass-weighted age was normalized by the Hubble time at each redshift, this relation would be flat. Overall, we find that FRB hosts at all redshifts follow this field galaxy trend.

\subsection{The Host in the Context of Compact Group Galaxies and Progenitor Implications}
\label{sec:disc-cggal-comp}

The discovery of FRB\,20220610A in a compact galaxy group marks a novel environment for FRBs. As highly dense structures, CGs are ideal sites for ongoing mergers and rapid galaxy interactions \citep{Mamon92}. The frequency of CGs at $z \approx 1$ is predicted to be more abundant than at $z \approx 0$ \citep{hp20}. However, CG galaxies comprise $\lesssim 1\%$ of all galaxies out to at least $z\approx 2$ \citep{wiens_etal_2019,dg20}, making this a fairly rare and noteworthy host environment. The large majority of observational CG studies are at low redshifts of $z \lesssim 0.2$. These studies find that CG galaxies generally exhibit redder colors, older stellar population ages, and are more morphologically compact compared to systems in less dense environments such as field galaxies \citep{Deng_etal_2008,Coenda_etal_2012,sohn2015,Montaguth_etal_2023}, seemingly at odds with a young stellar progenitor. However, at higher redshifts (toward $z \approx 2$), galaxies are more gas-rich and tidal interactions between galaxies can trigger bursts of star formation \citep{pearson_etal_2019}. Indeed, we find a ``bridge'' of diffuse material between the host of FRB\,20220610A and other galaxies within the group (e.g., G4 and G5) in both the F606W and F160W filters (Fig.~\ref{fig:imaging}, although only present in the F606W residual, denoted in Fig.~\ref{fig:galfit}), possibly indicative of a past or ongoing merger. This is commensurate with the large majority of CGs having at least one interacting pair \citep{mh94}. Thus, the existence of FRB\,20220610A in a compact group with signs of interaction could indicate its progenitor was associated with a fairly recent, young population of stars.

\subsection{A Comparison to Non-Repeating and Galaxy Cluster FRB Hosts} \label{sec:disc-frbhost-comp}

As FRB\,20220610A is an apparent non-repeater, we next investigate how it compares to the hosts of other non-repeaters. While no statistically significant differences have been found between the host galaxies of repeaters and apparent non-repeaters, the latter tend to occur in more massive and luminous galaxies than repeaters \citep{Bhandari+22,Gordon+23}. The fraction of transitioning and quiescent non-repeater host galaxies also appears to be larger than repeaters, with all quiescent hosts and the majority of transitioning hosts to date associated with non-repeaters \citep{Gordon+23, Sharma_etal_2023, Law+23}. However, these less active galaxies comprise a minority of the non-repeating (and thus overall) FRB host population; indeed, the large majority of non-repeating host galaxies are star-forming, aligning with the host of FRB\,20220610A.

\begin{figure}
    \centering
    \includegraphics[width=0.5\textwidth]{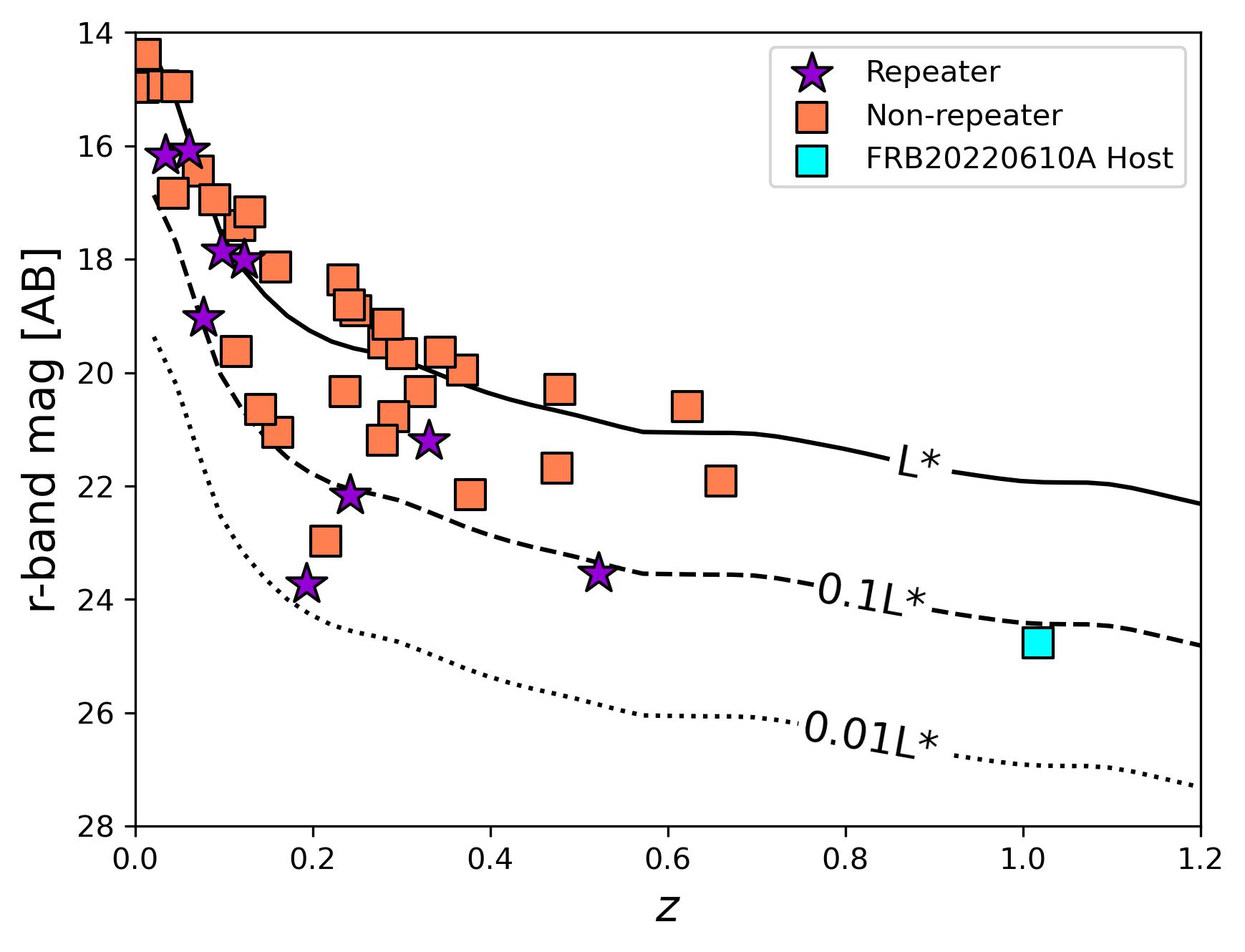}
    \caption{$r$-band AB magnitude versus redshift for published FRB host galaxies with available data (\citealt{Ravi+19,Gordon+23,Ibik+23,LeeWaddell+23,Law+23,Panther+23,Ravi23}).
    We represent repeaters with purple stars, non-repeaters with orange squares, and FRB\,20220610A with a cyan square. We plot the tracks of $L^*$, 0.1\,$L^*$, and 0.01\,$L^*$ galaxies in this parameter space. While the majority of $z\lesssim0.5$ FRB hosts trace $L^*$ galaxies, the host of FRB\,20220610A traces 0.1\,$L^*$ galaxies.}
    \label{fig:L*-curves}
\end{figure}

To compare the host of FRB\,20220610A to the larger FRB host population, we compare their optical luminosities with the redshift-evolving galaxy luminosity function (characterized by the characteristic luminosity $L^*$ in a Schechter galaxy luminosity function; \citealt{Schechter}). In Figure~\ref{fig:L*-curves}, we plot the $r$-band or $R$-band apparent AB magnitude versus redshift for the host of FRB\,20220610A and 42 FRB hosts with such data  \citep{Ravi+19,Gordon+23,Ibik+23,LeeWaddell+23,Law+23,Panther+23,Ravi23}. In the single case of FRB\,20190102C, we use $I$-band as $r/R$-band data were not available. We then plot curves corresponding to $L^*$, 0.1\,$L^*$, and 0.01\,$L^*$ galaxies, in which the values evolve with redshift but always correspond to the rest-frame $r$-band \citep{Brown+01,Wolf+03,Willmer+06,Reddy+09,Finkelstein+15,Heintz+20}. Nearly all non-repeating FRB hosts are $\gtrsim0.1\,L^*$, with the majority close to the $L^*$ curve. The lack of observed fainter hosts could be intrinsic to the population, or possibly an observational bias given that it is easier to detect bright galaxies and subsequently associate them to FRBs. The host of FRB\,20220610A is consistent with a luminosity of $\sim0.1$\,$L^*$, which is on the faint end of the observed luminosity function for non-repeating FRB hosts, but still consistent with the larger non-repeater population.

To date, FRB\,20220610A is the fourth FRB associated to a dense galactic environment. It joins FRBs\,20220509G and 20220914A which were discovered by the Deep Synoptic Array (DSA-110; \citealt{Ravi23}) and localized to hosts in the galaxy clusters Abell 2311 and 2310, respectively \citep{Connor_etal_2023,Sharma_etal_2023}, and FRB\,20200320A, a CHIME candidate repeater in the vicinity of several WISE$\times$SCOS galaxies but with a large localization region \citep{Rafiei_Ravandi_etal_2023}. As these FRBs occurred in a much lower redshift regime (with all three at $z\lesssim0.3$), we cannot compare their stellar population properties directly, but we note several interesting similarities. First, all three are apparent non-repeaters. Second, two of the three (FRB\,20220610A and FRB\,20220914A) are actively star-forming, tracing the star-forming main sequence when compared to other galaxies at their redshift. As galaxies within clusters and groups tend to be redder and more quiescent than field galaxies especially at low redshifts, \citep{Balogh+98,Balogh+04,Coenda_etal_2012}, it is notable that these FRBs occurred in star-forming galaxies, seemingly irrespective of the trends of the large-scale environment in which the galaxies reside. However, FRB\,20220509G was localized to a quiescent galaxy, indicating a diversity of global environments within this sub-population. 

While the fraction of FRBs confirmed to originate from galaxy clusters and groups is small (three out of $\approx50$ FRB hosts to date; this work, \citealt{Connor_etal_2023,Sharma_etal_2023}), it would be notable if this fraction increases as the number of localized hosts increases, especially at high redshift where the fraction of star-forming galaxies is higher \citep{Whitaker12}. Galaxy clusters are much more prevalent at low redshift, as these structures take time to form \citep{Springel+18}, so it is not surprising that the two cluster associations are at low-redshift. However, observational biases may dissuade association with such environments at all redshifts. First, the gas between the galaxies can contribute to the DM (Section~\ref{sec:disc-FRB-comp}), and there is more limited sensitivity of experiments to high-DM FRBs (i.e., Figure 4 of \citealt{Ryder23}). Second, thus far, only non-repeating FRBs have been discovered in these environments; if this trend continues, the absence of repeats may make it more difficult to reliably obtain sub-arcsecond localizations. Finally, at higher redshifts, it is clear that the resolution of {\it HST} is required to resolve close galaxy systems or pairs, which will be an observational challenge as the number of $z\gtrsim 1$ FRBs increases. While these associations to dense, galactic environments may be more difficult, FRBs in these environments act as unique probes of the intragroup (IGrM) or intracluster (ICM) media. Additional associations will also be useful to understand if large-scale environment plays a role in the FRB progenitor.

\subsection{The Effect of a Compact Galaxy Group Origin on FRB Properties} \label{sec:disc-FRB-comp}

We next investigate the likely effect of a compact galaxy group origin on FRB properties, and comment on if this is reflected in the observed properties of FRB\,20220610A. The IGrM is expected to be a multi-phase medium composed of very hot ($\sim10^{7}$~K), warm ($\sim10^{5} - 10^{6}$~K), and cool ($\leq10^{5}$~K) gas \citep{Oppenheimer+21}. It cools primarily via line emission as opposed to Bremsstrahlung radiation as in the ICM \citep{Lovisari+21,Oppenheimer+21}. Given that it is ionized, it should contribute to the observed FRB DM in a similar manner to the ICM in galaxy clusters \citep{xyz19}. If it also contains a significant ordered magnetic field, this ionized medium would contribute to the observed RM. Finally, if the IGrM is sufficiently turbulent, it would additionally contribute to multi-path propagation that could be observed via temporal broadening of the FRB pulse (i.e., scattering), scintillation, or depolarization. We treat each of these possibilities in turn in the following subsections.

\subsubsection{Can Intragroup Gas Account for the Excess DM?} \label{sec:disc-DMexcess}

\begin{figure*}[t]
    \centering
    \includegraphics[width=\textwidth]{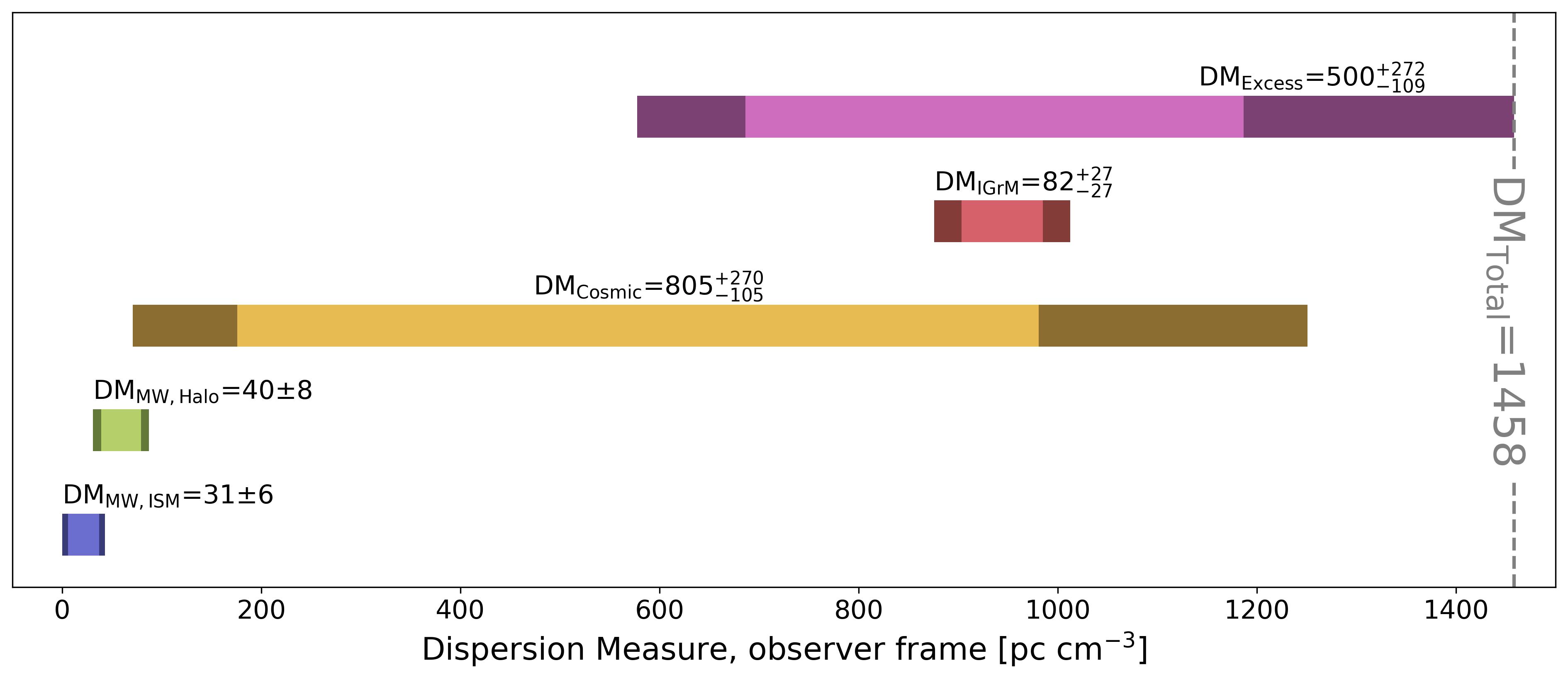}
    \caption{The DM budget of FRB\,20220610A composed of $\rm{DM}_{\rm{MW,ISM}}$ (blue), $\rm{DM}_{\rm{MW,Halo}}$ (green), $\rm{DM}_{\rm{cosmic}}$ (yellow), $\rm{DM}_{\rm IGrM}/(1+z_{\rm IGrM})$ (red), and $\rm{DM}_{\rm host}/(1+z_{\rm host})$ (pink), all in the observed frame. We assume 20\% uncertainties on the MW ISM and halo components \citep{Simha+23}, and take $z_{\rm IGrM} = z_{\rm host} = 1.017$. We find $\rm{DM}_{\rm IGrM}/(1+z_{\rm IGrM})$ can account for $82\pm27~\dmunits$, leaving an excess DM of 500$^{+272}_{-109}$ \dmunits.}
    \label{fig:DM-budget}
\end{figure*}

The total observed DM of an FRB (\dmfrb) is made up of contributions from the Milky Way, the cosmic web and intervening galaxy halos, and the host galaxy, i.e., 

\begin{equation}
\dmfrb = \rm{DM}_{\rm{MW}} + \rm{DM}_{\rm{cosmic}} + \frac{\rm{DM}_{\rm{host}}}{1+z_{host}}.  
\end{equation}
$\rm{DM}_{\rm{MW}} = \rm{DM}_{\rm{MW,ISM}} + \rm{DM}_{\rm{MW,halo}}$ can be estimated from maps of the electron distribution in the Milky Way and assumptions on the Milky Way dark matter halo. $\rm{DM}_{\rm{cosmic}}$ encompasses gas in the circumgalactic medium (CGM) from potential intervening halos and the diffuse intergalactic
medium (IGM); the mean value can be estimated via the Macquart relation \citep{mpm+20}. The remaining DM is attributed to ionized material in and around the host galaxy via $\rm{DM}_{\rm{host}}/(1+z_{host})$.  While the size of the $\rm{DM}_{\rm{host}}$ contribution must depend on the nature and orientation of the host galaxy, the location of the FRB within it, and the presence of any dense circumburst medium, studies to date have often assumed that $\rm{DM}_{\rm{host}}$ can be approximated by a constant value \citep[e.g.][]{Macquart+10} or drawn from a single distribution \citep[e.g.][]{James+22}. Furthermore, if the FRB host is located in a galaxy cluster or group, the corresponding ICM and/or IGrM contributions would also contribute to the $\rm{DM}_{\rm{host}}/(1+z_{\rm host})$ term. 

There have been several FRBs whose extragalactic DMs (i.e., all components outside of the Milky Way) are anomalously high compared to the value of $\rm{DM}_{\rm{cosmic}}$ inferred from the Macquart relation; these are often referred to as ``DM-excess'' sources (e.g., \citealt{Niu+22,Simha+23}). Indeed, FRB 20220610A is one of these ``DM-excess'' FRBs. One possible explanation for these sources is a larger than assumed value for $\rm{DM}_{\rm{host}}$. Indeed, there is precedent for attributing excess DM to the ICM in the the case of FRB\,20220914A, which was localized to a galaxy cluster. \citet{Connor_etal_2023} attribute most of the excess DM to the ICM. However, the ICM DM contribution is larger than any expected DM contribution from the IGrM owed to the much larger masses of galaxy clusters over groups. 

For FRB\,20220610A, \citet{Ryder23} measured a $\dmfrb$ of 1458 \dmunits.\ Assuming $\rm{DM}_{\rm MW}= 71\ \dmunits$ (with $\rm DM_{\rm MW,ISM} = 31\ \dmunits$ from NE2001 \citep{NE2001} and $\rm{DM}_{\rm MW,halo} = 40\ \dmunits$; \citealt{xyz19}) and an average $\rm{DM}_{\rm cosmic} = 805\ \dmunits$ at a redshift of $z=1.017$, this sightline is in excess by $582\ \dmunits$. If this arose entirely from the host galaxy (i.e., $\rm{DM}_{\rm{host}} = 1173\ \dmunits$ in the rest frame), FRB\,20220610A would have the most extreme host DM contribution to date, a factor of two higher than the next highest value (for FRB\,20190520B; \citealt{Simha+23, Lee+23}).

With FRB\,20220610A now confirmed to reside in a compact group of galaxies, the IGrM could potentially provide a natural explanation for at least some of the excess DM. To estimate \dmigrm, we must first make assumptions on the spatial distribution of the ionized gas in the IGrM. Since our spectroscopic data is limited in resolution and does not encompass every galaxy in the group, we cannot estimate the velocity dispersion of the group members to infer the group virial mass. Instead, we derive a rough estimate of \dmigrm\ by assuming a dark matter halo mass based on the combined stellar masses of the galaxies. We first estimate the stellar masses for all galaxies using the \citet{van_der_Wel+14} mass-radius relation, assuming $z=1.017$ and the \textsc{Galfit} effective radii measurements in the F160W filter. Using the methods described in \citet{Nugent+23}, we take the highest likelihood mass as the stellar mass of the galaxy; these values are reported in Table~\ref{tab:host/gal_props}. For G1 and G6, we instead use the \texttt{Prospector}-derived stellar mass estimates as described in Section~\ref{sec:SPP-modeling}. Next, we use the abundance-matching based stellar-to-halo-mass ratio (SHMR) from \citet{moster+13} to convert the stellar masses of the galaxies to halo mass estimates. We assume the group halo mass is a simple sum of the individual halo mass estimates, i.e., $\approx10^{12} M_\odot$. Then, following \citet{Simha+23} and \citet{Lee+23}, we assume the plasma in the IGrM follows a modified NFW profile with $y_0=2$ and $\alpha=2$ which extends to 2 virial radii \citep[see][for a mathematical description of the model]{xyz19}. We further assume the ionized fraction of the baryons within the group, $f_{\rm gas}$, is 0.6. Our choices for these model parameters are justified by the work of \citet{Khrykin+2023} who investigated the partition of baryons in halos and the diffuse IGM. 

We estimate a fiducial contribution of \dmigrm\ to the observed FRB DM by assuming the FRB is at the center of the halo gas profile. This results in $\dmigrm \approx 55~\dmunits$ in the observer frame and $110~\dmunits$ in the rest frame. If the FRB were located at the far outer edge of the halo and intersected the halo diameter, at most, $\dmigrm$ could contribute $\approx 110~\dmunits$, or $220\ \dmunits$ in the rest frame. If instead the FRB was offset from the center, the sightline would intersect a smaller portion of the halo and yield a commensurately lower $\dmigrm$. Thus, we assume a fiducial contribution of $\dmigrm = 82\pm27~\dmunits$ to the IGrM (or $165 \pm 55$\ \dmunits\ in the rest-frame). We calculate the resulting excess DM as $500^{+272}_{-109}$ \dmunits\ in the observed frame (or $1008^{+548}_{-220}$ in the rest-frame), estimating the uncertainties via Gaussian error propagation on all preceding DM components. We provide a visual for the relative contributions of these components in Figure~\ref{fig:DM-budget}.  

There are two possible scenarios to explain the remaining excess DM: (1) The majority of the excess DM must be accounted for by the remaining components of DM$_{\rm host}$, i.e., the circumburst environment and/or the host ISM; or (2) there are significant foreground structures along the sightline, meaning DM$_{\rm cosmic}$ is much greater than the average value inferred from the Macquart relation (as was the case for FRB~20190520B; \citealp{Lee+23}).\footnote{We note a combination of these scenarios is additionally plausible.} Unless a comprehensive assessment of the foreground line-of-sight \citep[e.g.][]{Simha+23, Lee+23} indicates an unusually dense sightline to FRB\,20220610A, the substantial excess DM must reside within the host galaxy. Thus, further follow-up with line-of-sight foreground mapping will be required to disentangle these scenarios and place constraints on the properties of, and contributions from, the IGrM. At the same time, a more robust analysis of the \dmigrm\ could be performed with redshifts of all the galaxies in the system. Space-based IFU observations with e.g., \textit{JWST}, would be a promising pathway for more detailed analysis.

\subsubsection{Revisiting the Rotation Measure} \label{sec:disc-RM}

The RM can be used to probe the magneto-ionic environment of the FRB, and when combined with the DM, can place constraints on the magnetic fields of the host galaxy. The now-measured presence of an IGrM (which alters the DM budget for the host ISM and circumburst medium) invites the RM budget of FRB\,20220610A to be revisited.

As the IGM is not expected to contribute significantly to the RM \citep{Mannings+23}, the measured RM of an FRB (neglecting the Earth's ionospheric contribution of a few rad\,m$^{-2}$ and the small MW contribution) can be attributed mainly to the host galaxy. However, in the case of intervening galaxy clusters or groups (either line-of-sight or local to the FRB), the ICM or IGrM could contribute measurably to the observed RM \citep{Connor_etal_2023,Sherman+23}.

The rest-frame RM$_{\rm host}$ of FRB\,20220610A is 860 rad~m$^{-2}$ \citep{Ryder23}, a factor of five higher than the median rest-frame $|$RM$_{\rm host}|$ of the non-repeating FRB population (163.91 rad~m$^{-2}$; data from \citealt{Sherman+23,Mannings+23}) and a factor of 36 higher than the median rest-frame $|$RM$_{\rm host}|$ of the repeating FRB population\footnote{We note the range of the repeating FRB RMs is significantly larger than that of non-repeaters due to the extreme RM of  FRB\,20121102A, which, while varying, has been measured to be as large as 1.6$\times10^{5}$ rad~m$^{-2}$ \citep{2021ApJ...908L..10H}.} (24.2 rad~m$^{-2}$; data from \citealt{Mannings+23}). When combined, the DM and RM probe the magnetic field along the line-of-sight via

\begin{equation}
    B_\parallel = 1.23~ \mu {\rm G}~ \frac{\rm RM}{\rm DM}.
\end{equation}

\noindent If the RM originates in plasma at the redshift of the burst host (either in the circumburst medium, the host galaxy ISM, or in the IGrM), this would correspond to a (electron density weighted) mean magnetic field strength parallel to the line of sight of $B_\parallel=$ 0.902 $\mu$G, assuming a DM$_{\rm excess+IGrM}$ of 1173 pc cm$^{-3}$. This magnetic field strength is typical for pulsar lines of sight through the Milky Way ISM \citep{Ryder23}, but is larger than expected for the diffuse IGrM (e.g., \citealt{Prochaska+19} calculate an upper limit of $<0.8 \mu$G for the host halo of FRB\,20181112A). Combined with the fact that the IGrM contributes only a minority of the DM excess, we conclude that it could only contribute a \textit{majority} of the RM if the $B_\parallel$ in the IGrM was considerably larger than expectations. Without an independent constraint on RM$_{\rm IGrM}$, we cannot further separate $B_{\parallel,\rm host}$ and $B_{\parallel,\rm IGrM}$, but it is reasonable to assume that while both locations contribute to the total $B_\parallel$, the host ISM dominates.

\subsubsection{Multi-path Propagation} \label{sec:disc-mpp}

Given the new observations, we can also revisit the origin of the scatter broadening of the FRB. \citet{Ryder23} estimated the scatter broadening of FRB\,20220610A to be $1.88 \pm 0.02$\,ms. For an equivalent column of dispersion, the strength of the scatter broadening is approximately three orders of magnitude lower than that of the Milky Way ISM (\citealt{2004ApJ...605..759B}; after correcting for the redshift of the scattering plasma, differences in geometry between Milky Way lines of sight, and the ISM of the host galaxy. Thus, comparison of the pulse scattering to the inferred electron density favors plasma close to the FRB progenitor as the source of the scattering.

While $\rm{DM}_{\rm IGrM}$ is a minority of the total inferred $\rm{DM}_{\rm host}$, the geometrical lever arm effect of locating it further from the FRB progenitor means that if this plasma was equally turbulent to a typical ISM sightline, it could nevertheless dominate the observed scattering. However, observations of other FRBs which have probed halo sightlines inferred very low levels of turbulence which would contribute negligibly to the observed scattering \citep{Prochaska+19}. If, however, the IGrM does contribute appreciably to the measured scattering, this would further strengthen the argument that dense plasma in the host galaxy must reside close to the FRB progenitor, as has been inferred for other FRB systems \citep{2023MNRAS.519..821O}.

In summary, if subsequent observations show no overdensity of foreground structures along the line of sight and a large excess DM {\em is} associated with the host galaxy, it likely originates in plasma close to the FRB progenitor. However, this plasma only produces moderate scattering and Faraday rotation, and does not exhibit frequency-dependent depolarization \citep{Ryder23}. Taken together, these properties are consistent with a dense circumburst medium similar to those inferred for repeating FRBs that exhibit DM excesses, but less magneto-ionically active than inferred for those sources. Alternatively, the host galaxy ISM could be (very) significantly less turbulent than the Milky Way's ISM. Of course, if the $\rm{DM}_{\rm host}$ budget is significantly reduced because $\rm{DM}_{\rm cosmic}$ is found to be higher due to foreground structures, the dominant locations of the scattering (and Faraday rotation) would have to be revisited.

\section{Conclusions} \label{sec:conc}

We have presented two-filter \textit{HST} imaging of the host of FRB\,20220610A, the FRB with the most distant confirmed host to date. These observations reveal a previously unresolved complex composed of at least seven sources within a region of radius 16~kpc, two of which were too faint to be detected in ground-based imaging. In the context of the {\it HST} imaging, we re-visit previous VLT spectroscopy and determine spectroscopic redshifts for at least four of the galaxies at $z \approx 1$. Owing to their similar redshifts and the lack of clear evidence for a larger-scale galaxy cluster, we identify the galaxies as members of a compact group. We come to the following conclusions:

\begin{itemize}
    \item We identify the galaxy coincident with the FRB, G1, as the most likely host of FRB\,20220610A. The next most likely host galaxy (G6) is a factor of $\sim 8$ less probable. We measure the redshift of the host to be $z=1.017$.
    \item We use SED modeling to determine the stellar population properties of the host galaxy and derive a stellar mass of log(M$_*$/M$_{\odot}$) $=$ 9.69$\pm$0.11, a mass-weighted age of $t_{\rm m}$ $=2.60^{+0.61}_{-0.91}$~Gyr, and a star formation rate integrated over the past 100~Myr of ${\rm SFR}_{\rm 0-100~Myr}$ $=1.67^{+2.41}_{-0.95}$~M$_{\odot}$~yr$^{-1}$. These properties designate the host as a star-forming galaxy.
    \item The host galaxy properties are consistent with the star-forming field galaxy population at $z\sim 1$ in stellar mass, star-formation rate, star formation history, and mass-weighted age. However, the host is slightly more massive than star-forming field galaxies at its redshift.
    \item Comparing the host of FRB\,20220610A to the hosts of non-repeating FRBs (the large majority of which reside at $z \lesssim 0.5$), we find that it traces field galaxy properties analogously to the low-$z$ FRB host population. However, while most low-$z$ hosts have luminosities around the characteristic luminosity of the galaxy luminosity function ($L^*$), the host of FRB\,20220610A is on the fainter end of the luminosity function at $\sim0.1\,L^*$. 
    \item The {\it HST} imaging reveals potential signs of interaction between the host and other galaxy group members, which are thought to trigger bursts of star formation in these physically dense systems. Coupled with its existence in a star-forming galaxy, this could indicate that the progenitor of FRB\,20220610A is associated with a fairly recent population of stars.  
    \item In light of the compact group origin, we examine the intragroup medium as the source of the excess DM. Our fiducial estimate of \dmigrm\ can account for $82\pm27~\dmunits$ in the observed frame (165$\pm$55~\dmunits\ in the rest-frame). Thus, we conclude the IGrM is a minority contributor to the host DM, leaving an excess DM of $500^{+272}_{-109}$ \dmunits\ in the observed frame (1008$^{+548}_{-220}$~\dmunits\ in the rest-frame). 
    \item The excess DM likely originates from either (1) the immediate circumburst environment to the FRB progenitor and/or (2) significant foreground structures along the line of sight, or (3) a combination of the two. Further follow-up with line-of-sight foreground mapping will be required to disentangle these scenarios. 
    \item The magnetic fields along the line of sight are unlikely to be dominated by the IGrM but rather by the host ISM. However, without an independent constraint on RM$_{\rm IGrM}$, we cannot further separate the magnetic field contributions of the IGrM and the host galaxy. 
\end{itemize}

\noindent FRB\,20220610A marks the fourth FRB associated with a dense, large-scale environment, and is the first at $z \gtrsim 0.3$, providing a benchmark event for the field. Furthermore, it is the first confirmed $z\approx1$ FRB to be localized to its host galaxy. Upgrades to current FRB detection facilities and new experiments have been coming online that are increasing the sensitivity to fainter, high-DM FRBs. Thus, we expect the $z\gtrsim1$ FRB host population to increase steadily over the next few years. This population will provide crucial context for the host of FRB\,20220610A and enable studies of the evolution of FRB and host properties over cosmic time by bridging this population to those at low-$z$. To provide an unambiguous association to their host galaxies, the localizations of $z\gtrsim1$ FRBs must be $\lesssim0\farcs5$. Larger samples of FRB host galaxies will be required to quantify the frequency of sub-populations, such as those associated with dense, large-scale environments, and the resulting implications for the FRB progenitor(s), their formation channels, and their usage as probes of the cosmic web. This is especially crucial for hosts that are not immediately detected or easily characterized in existing archival imaging, as these ``outliers'' can place strong constraints on the nature of the FRB progenitor(s).

\section{Acknowledgments}
The authors thank Mitchell Revalski for guidance on PSF modeling, Clecio Bom and Gabriel Teixeira for sharing DELVE photometric redshifts, Stephanie Juneau for discussion on field galaxy photometric redshifts, Daniele Sorini for his calculation of $f_{\rm gas}$ in halos at z = 1, Khee-Gan Lee for discussions on the DM properties of the IGrM, Gary Mamon for insightful discussions on compact galaxy groups, and Kameswara Bharadwaj Mantha for contributions to the \textit{HST} proposal.

A.C.G. and the Fong Group at Northwestern acknowledges support by the National Science Foundation under grant Nos. AST-1814782, AST-1909358 and CAREER grant No. AST-2047919. W.F. gratefully acknowledges support by the David and Lucile Packard Foundation, the Alfred P. Sloan Foundation, and the Research Corporation for Science Advancement through Cottrell Scholar Award \#28284. A.C.G., W.F., S.S., Y.D., C.D.K., T.E., A.R.M., J.X.P., and N.T. acknowledge support from NSF grants AST-1911140, AST-1910471 and AST-2206490 as members of the Fast and Fortunate for FRB Follow-up team. R.M.S. acknowledges support through Australian Research Council Future Fellowship FT190100155. R.M.S and A.T.D acknowledge support through Australian Research Council Discovery Project DP220102305. C.D.K. acknowledges partial support from a CIERA postdoctoral fellowship. L.M. acknowledges the receipt of an MQ-RES scholarship from Macquarie University. Y.D. is supported by the National Science Foundation Graduate Research Fellowship under Grant No.421 DGE-1842165. T.E. is supported by NASA through the NASA Hubble Fellowship grant HST-HF2-51504.001-A awarded by the Space Telescope Science Institute, which is operated by the Association of Universities for Research in Astronomy, Inc., for NASA, under contract NAS5-26555. M.G. is supported by the Australian Government through the Australian Research Council's Discovery Projects funding scheme (DP210102103).

This research was supported in part through the computational resources and staff contributions provided for the Quest high performance computing facility at Northwestern University which is jointly supported by the Office of the Provost, the Office for Research, and Northwestern University Information Technology.

This scientific work uses data obtained from Inyarrimanha Ilgari Bundara/the Murchison Radio-astronomy Observatory. We acknowledge the Wajarri Yamaji People as the Traditional Owners and native title holders of the Observatory site. CSIRO’s ASKAP radio telescope is part of the Australia Telescope National Facility (https://ror.org/05qajvd42). Operation of ASKAP is funded by the Australian Government with support from the National Collaborative Research Infrastructure Strategy. ASKAP uses the resources of the Pawsey Supercomputing Research Centre. Establishment of ASKAP, Inyarrimanha Ilgari Bundara, the CSIRO Murchison Radio-astronomy Observatory and the Pawsey Supercomputing Research Centre are initiatives of the Australian Government, with support from the Government of Western Australia and the Science and Industry Endowment Fund.

Based on observations collected at the European Organisation for Astronomical Research in the Southern Hemisphere under ESO programmes 105.204W and 108.21ZF.

The Legacy Surveys consist of three individual and complementary projects: the Dark Energy Camera Legacy Survey (DECaLS; Proposal ID \#2014B-0404; PIs: David Schlegel and Arjun Dey), the Beijing-Arizona Sky Survey (BASS; NOAO Prop. ID \#2015A-0801; PIs: Zhou Xu and Xiaohui Fan), and the Mayall z-band Legacy Survey (MzLS; Prop. ID \#2016A-0453; PI: Arjun Dey). DECaLS, BASS and MzLS together include data obtained, respectively, at the Blanco telescope, Cerro Tololo Inter-American Observatory, NSF’s NOIRLab; the Bok telescope, Steward Observatory, University of Arizona; and the Mayall telescope, Kitt Peak National Observatory, NOIRLab. Pipeline processing and analyses of the data were supported by NOIRLab and the Lawrence Berkeley National Laboratory (LBNL). The Legacy Surveys project is honored to be permitted to conduct astronomical research on Iolkam Du’ag (Kitt Peak), a mountain with particular significance to the Tohono O’odham Nation.

NOIRLab is operated by the Association of Universities for Research in Astronomy (AURA) under a cooperative agreement with the National Science Foundation. LBNL is managed by the Regents of the University of California under contract to the U.S. Department of Energy.

This project used data obtained with the Dark Energy Camera (DECam), which was constructed by the Dark Energy Survey (DES) collaboration. Funding for the DES Projects has been provided by the U.S. Department of Energy, the U.S. National Science Foundation, the Ministry of Science and Education of Spain, the Science and Technology Facilities Council of the United Kingdom, the Higher Education Funding Council for England, the National Center for Supercomputing Applications at the University of Illinois at Urbana-Champaign, the Kavli Institute of Cosmological Physics at the University of Chicago, Center for Cosmology and Astro-Particle Physics at the Ohio State University, the Mitchell Institute for Fundamental Physics and Astronomy at Texas A\&M University, Financiadora de Estudos e Projetos, Fundacao Carlos Chagas Filho de Amparo, Financiadora de Estudos e Projetos, Fundacao Carlos Chagas Filho de Amparo a Pesquisa do Estado do Rio de Janeiro, Conselho Nacional de Desenvolvimento Cientifico e Tecnologico and the Ministerio da Ciencia, Tecnologia e Inovacao, the Deutsche Forschungsgemeinschaft and the Collaborating Institutions in the Dark Energy Survey. The Collaborating Institutions are Argonne National Laboratory, the University of California at Santa Cruz, the University of Cambridge, Centro de Investigaciones Energeticas, Medioambientales y Tecnologicas-Madrid, the University of Chicago, University College London, the DES-Brazil Consortium, the University of Edinburgh, the Eidgenossische Technische Hochschule (ETH) Zurich, Fermi National Accelerator Laboratory, the University of Illinois at Urbana-Champaign, the Institut de Ciencies de l’Espai (IEEC/CSIC), the Institut de Fisica d’Altes Energies, Lawrence Berkeley National Laboratory, the Ludwig Maximilians Universitat Munchen and the associated Excellence Cluster Universe, the University of Michigan, NSF’s NOIRLab, the University of Nottingham, the Ohio State University, the University of Pennsylvania, the University of Portsmouth, SLAC National Accelerator Laboratory, Stanford University, the University of Sussex, and Texas A\&M University.

BASS is a key project of the Telescope Access Program (TAP), which has been funded by the National Astronomical Observatories of China, the Chinese Academy of Sciences (the Strategic Priority Research Program “The Emergence of Cosmological Structures” Grant \# XDB09000000), and the Special Fund for Astronomy from the Ministry of Finance. The BASS is also supported by the External Cooperation Program of Chinese Academy of Sciences (Grant \# 114A11KYSB20160057), and Chinese National Natural Science Foundation (Grant \# 12120101003, \# 11433005).

The Legacy Survey team makes use of data products from the Near-Earth Object Wide-field Infrared Survey Explorer (NEOWISE), which is a project of the Jet Propulsion Laboratory/California Institute of Technology. NEOWISE is funded by the National Aeronautics and Space Administration.

The Legacy Surveys imaging of the DESI footprint is supported by the Director, Office of Science, Office of High Energy Physics of the U.S. Department of Energy under Contract No. DE-AC02-05CH1123, by the National Energy Research Scientific Computing Center, a DOE Office of Science User Facility under the same contract; and by the U.S. National Science Foundation, Division of Astronomical Sciences under Contract No. AST-0950945 to NOAO.

This work has made use of data from the European Space Agency (ESA) mission
{\it Gaia} (\url{https://www.cosmos.esa.int/gaia}), processed by the {\it Gaia}
Data Processing and Analysis Consortium (DPAC,
\url{https://www.cosmos.esa.int/web/gaia/dpac/consortium}). Funding for the DPAC
has been provided by national institutions, in particular the institutions
participating in the {\it Gaia} Multilateral Agreement.

%

\vspace{5mm}
\facilities{\textit{HST} (WFC3),
Keck:II (DEIMOS),
VLT:Antu (FORS2),
VLT:Yepun (HAWK-I),
VLT:Kueyen (X-Shooter)
}


\software{
\texttt{astropy} \citep{astropy},
\texttt{drizzlepac} \citep{drizzlepac},
\texttt{dynesty} \citep{Speagle2020},
\texttt{FRBs/astropath}\footnote{https://github.com/FRBs/astropath},
\texttt{FRBs/FRB} \citep{F4_repo},
\texttt{FSPS} \citep{Conroy2009,Conroy2010},
\texttt{galfit} \cite{galfit_2002,galfit_2010},
\texttt{ginga}\footnote{https://github.com/ejeschke/ginga},
\texttt{hst\char`_wfc3\char`_psf\char`_modeling} \citep{PSF_zenodo,Revalski_etal_2023},
\texttt{matplotlib} \citep{matplotlib},
\texttt{numpy} \citep{numpy}, 
\texttt{pandas} \citep{pandas},
\texttt{photutils} \citep{photutils}, 
\texttt{Prospector} \citep{Johnson+21},
\texttt{python-FSPS} \citep{python-fsps},
\texttt{SAOImageDS9} \citep{DS9},
\texttt{scipy} \citep{scipy},
\texttt{seaborn} \citep{seaborn},
\texttt{sedpy} \citep{sedpy},
\texttt{Source Extractor} \citep{source_extractor}
}



\appendix

\renewcommand\thefigure{\thesection.\arabic{figure}}   
\setcounter{figure}{0} 

\section{Emission Line Identification}
\label{app:emission-lines}

\begin{figure*}[t]
    \centering   
    \includegraphics[width=\textwidth]{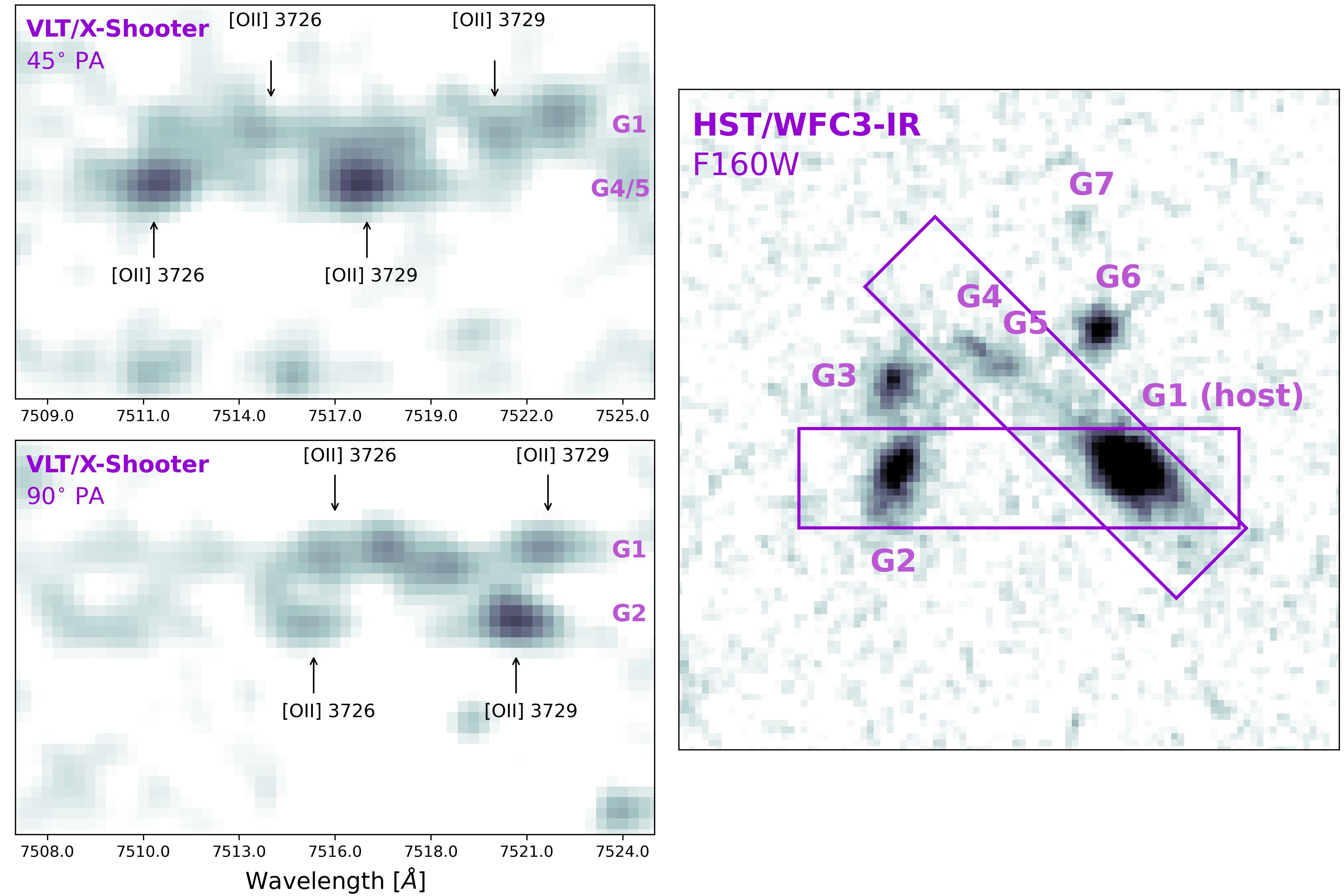}
    \caption{(Left:) 45$^{\circ}$ (top) and 90$^{\circ}$ (bottom) position angle X-Shooter spectra highlighting the [OII] 3726\AA/3729\AA~doublet at the positions of G1, G2, G4, and G5. The spectra are smoothed with a 1.5$\sigma$ Gaussian filter to highlight the emission features. (Right:) The F160W image with the X-Shooter slits overlaid. We note that in addition to the real [OII] doublet features present at the same locations in both spectra, there are also noise features in the spectrum of G1, identified as such because they do not appear in the same places across both slit position angles.}
    \label{fig:emission-line-fig}
\end{figure*}

\citet{Ryder23} obtained two spectra with VLT/X-Shooter (PI Macquart, Program 105.204W.004; \citealt{X-shooter}), one at a position angle of 45$^{\circ}$ and the second at 90$^{\circ}$, to determine if the objects identified in the ground-based imaging belonged to one galaxy or three (i.e., Figure 2 of \citealt{Ryder23}). Through detections of the [O\,II]$\lambda$3726\AA, 3729\AA~doublet and H$\alpha$, the redshift of the three clumps was established at $1.016\pm0.002$; one of the interpretations is that the clumps belonged to a single galaxy. 

As briefly described in Section~\ref{sec:system}, we overlaid the slit orientations on the \textit{HST} imaging to determine the redshifts of the galaxies that are now resolved. The spectrum at PA=90$^{\circ}$ covered the location of G1 and G2, and the PA=45$^{\circ}$ spectrum covered G1, G4, and G5. The centers of G1 and G2 are separated by 2\farcs17, which translates to a separation of 6.73~pixels at the pixel scale and binning of the X-Shooter data. We additionally use an unrelated object that serendipitously aligned with the slit to further verify the galaxy locations. We identify the [O\,II]$\lambda$3726\AA, 3729\AA~doublet at the expected location of G1 and G2, with redshifts of $z=1.017$. G1 and G4 are separated by 1\farcs74 (5.51~pixels); G1 and G5 are separated by 1\farcs47 (4.65~pixels). Via identification of the [O\,II] doublet, we again confirm G1 to be at $z=1.017$, while G4 and G5 are consistent with $z=1.016$ (they are not resolved into two components in the spectra). We show the 2D spectra centered on the [O\,II]$\lambda$3726\AA, 3729\AA~doublet as well as their positions on the sky overlaid on the F160W image in Figure~\ref{fig:emission-line-fig}. We obtained further follow-up spectroscopy to identify the redshift of G6 and find via probable identification of the [O\,II] doublet that the redshift is consistent with $z \approx 1$. The low signal-to-noise of the line identifications prevents further precision on the redshift of G6.

\bibliography{main}{}
\bibliographystyle{aasjournal}


\end{CJK*}
\end{document}